\documentclass[11pt]{article}
\usepackage[bottom]{footmisc}
\usepackage[T1]{fontenc} 
\usepackage{setspace,geometry}
\usepackage[utf8]{inputenc}
\usepackage[british,english,american]{babel}
\usepackage[backend=biber,style=authoryear, sorting=nyt]{biblatex}
\usepackage{color,xcolor,fancybox}  
\usepackage{pgfplots,tikz}
\usetikzlibrary{intersections}
\usetikzlibrary{shapes.geometric, arrows,calc}
\usetikzlibrary{decorations.markings}
\definecolor{myCB}{RGB}{255, 107, 135}
\definecolor{myGreen}{RGB}{34, 139, 34}
\usepackage[final]{pdfpages}
\usepackage{parskip}
\usepackage{authblk,graphicx,float,comment,dsfont,textcomp,csquotes,enumerate,array,pdflscape,enumitem}
\usepackage[linesnumbered,ruled]{algorithm2e}
\usepackage{amsmath,amsfonts,pifont,amssymb,amsthm,amsbsy,bbm,bm,mathrsfs,calligra}
\usepackage{arydshln, array, ctable, calrsfs, booktabs,subfig,adjustbox, multirow,threeparttable, makecell}

\usepackage{mathtools}
\usepackage{footnote}
\DeclareAutoCiteCommand{inline}{\textcite}{\textcites}

\usepackage[figurename=Figure,tablename=Table]{caption}
\usepackage{wrapfig}
\usepackage[detect-all]{siunitx}
\makesavenoteenv{tabular}
\makesavenoteenv{table}

\usepackage{hyperref}
\hypersetup{
	colorlinks=true, %set true if you want colored links
	linktoc=all,     %set to all if you want both sections and subsections linked
	linkcolor=blue,  %choose some color if you want links to stand out
	citecolor=blue,
	urlcolor=blue!30
}
\makeatletter \LetLtxMacro{\BHFN@Old@footnotemark}{\@footnotemark} \renewcommand*{\@footnotemark}{\refstepcounter{BackrefHyperFootnoteCounter}\xdef\BackrefFootnoteTag{bhfn:\theBackrefHyperFootnoteCounter}\label{\BackrefFootnoteTag}\BHFN@Old@footnotemark} \makeatother

\makeatletter

\newcommand{\distas}[1]{\mathbin{\overset{#1}{\kern\z@\sim}}}%
\newsavebox{\mybox}\newsavebox{\mysim}
\newcommand{\distras}[1]{%
	\savebox{\mybox}{\hbox{\kern3pt$\scriptstyle#1$\kern3pt}}%
	\savebox{\mysim}{\hbox{$\sim$}}%
	\mathbin{\overset{#1}{\kern\z@\resizebox{\wd\mybox}{\ht\mysim}{$\sim$}}}%
}
\makeatother
\makeatletter
\renewcommand{\@algocf@capt@plain}{above}% formerly {bottom}
\makeatother

\newcommand{\mE}{\mathbb{E}}

\newcommand{\vx}{\mathbf{x}}
\newcommand{\vy}{\mathbf{y}}
\newcommand{\bep}{\bm \varepsilon}
\newcommand{\bmu}{\bm \mu}
\DeclareMathOperator*{\argmin}{arg\,min}
\usepackage{footnotebackref}
\usepackage{tablefootnote}
\usepackage{scalerel}
\usepackage{float}
\restylefloat{table}
\usepackage[none]{hyphenat}
\title{Forecasting high-frequency financial time series: \\ an adaptive learning approach with the order book data}
\author{Parley Ruogu Yang\footnote{Faculty of Mathematics, University of Cambridge, and Department of Statistics, University of Oxford. \\ Contact: ry266@cam.ac.uk  } }
\date{11 Sep 2020}% \\{\textbf{ VERSION: 1.0}} }
\begin{document}
%\includepdf[pages=-,width=1.1\textwidth]{../CS/CS.pdf}

\maketitle

\begin{center}
	\textbf{ABSTRACT}
\end{center}
	This paper proposes a forecast-centric adaptive learning model that engages with the past studies on the order book and high-frequency data, with applications to hypothesis testing.  In line with the past literature, we produce brackets of summaries of statistics from the high-frequency bid and ask data in the CSI 300 Index Futures market and aim to forecast the one-step-ahead prices. Traditional time series issues, e.g. ARIMA order selection, stationarity, together with potential financial applications are covered in the exploratory data analysis, which pave paths to the adaptive learning model. By designing and running the learning model, we found it to perform well compared to the top fixed models, and some could improve the forecasting accuracy by being more stable and resilient to non-stationarity. Applications to hypothesis testing are shown with a rolling window, and further potential applications to finance and statistics are outlined.


\vspace{1cm}

\begin{center}
	\textbf{ACKNOWLEDGEMENT
}\end{center}
I thank Dr Mihai Cucuringu (Department of Statistics, University of Oxford and The Alan Turing Institute) and Dr Alex Shestopaloff  (School of Mathematical Sciences, Queen Mary University of London and The Alan Turing Institute) for their advice and  support during the research. I thank CIFCO Guangzhou for providing the high-frequency dataset for which the empirical study can be based upon, and I am also grateful to St Anne’s College, University of Oxford for its Graduate Student Research Grant, for which the cost of running computing machine is partially funded.
\thispagestyle{empty}
\pagebreak

\thispagestyle{empty}
\tableofcontents
\pagebreak
\setcounter{page}{1}

	\pagebreak
\setcounter{page}{1}
\section{Introduction and Literature Review}
Time series can be described as a sequence of observations indexed by the time, which, by the nature of it, can be separated into the past and the future. The study of predicting the future based on the past information is defined as forecasting, which is of great importance to the society --- forecasting financial time series, e.g. the price of an asset, can be influential to the decisions of both the private and the public sectors.
As the computerisation of financial markets develops, higher frequency of the observation on the variables are taken and can be analysed. Consequently, forecasting such a high-frequency object becomes increasingly important. 

At a higher level, the statistical approach undertaken for learning the big data and obtaining better prediction has  evolved in the recent decades, in both the theory \parencite{SS2014} and the applications, e.g. LSTM and deep learning \parencites{LSTM}{DL}. Various methods of learning, e.g. clustering, neural networks and other synthetic models have been developed and many of which have helped to solve socio-economic problems \parencites{TS_Clust}{ChakrabortyJoseph}. The application of statistical learning algorithms to dynamically assess forecasts has also shown a contribution to the empirical time series econometrics literature \parencite{py}.

However, due to the nature that time series dataset is indexed by time, and the fact that many structures (mathematically, such a concept is quantified by functional forms and parameters) change over time, one needs to pay particular attention while applying  generically-developed learning methods to a financial time series environment \parencite{SirignanoCont}. Econometricians refer such a unique time series issue as "time-varying parameters", which could also relate to the stationarity of a model --- essentially questioning the validity of the boundedness of the variables over time.\footnote{One may refer to \textcite{Klenke2013} for a more rigorous definition on stationarity.} Recent proposals on dealing with these have been suggested by \textcite{AndresHarvey} and
\textcite{ACH2013}, with some empirical studies being done \parencite{ACH2014}. Additionally, adapting learning methods to improve traditional ARIMA models' forecast has also been studied empirically \parencite{LHS2020}. In this paper, we use the classical approach of window-estimation, thereby focusing on the contemporary relationship between variables to ensure time-variability. The choice of window sizes vary, as shown later in the adaptive learning, different sizes could be preferred from time to time.

In terms of generating features (explanatory variables) to help to forecast the price, the order book data becomes particularly helpful. In this paper, the order book data consists of the quantities and prices for the best bid and ask --- meaning the ones at which the asset can be traded immediately sold and bought, respectively. Statistics of these can be summarised into order flow imbalance, which synthetically involves the prices and quantities on both sides, or order imbalance, which deals with solely the quantities on both sides. The earlier has been studied by  \textcites{RSR2010}{CKS} and the latter by \textcites{ARS2011}{Stoikov}. 

While more sophisticated learning models could be used, e.g. deep learning models \parencite{SirignanoCont}, we start from a traditional time series modelling and statistical learning perspective and subsequently propose learning models that can adapt to the past (thus called adaptive learning). Such a model has a better interpretability and numerous potential applications, e.g. hypothesis testing. A general reference on the foundation of model selection is from \textcites{Akaike1974}{ESL2001}. 
Additionally, penalisation plays a key role in the study of time series model selection \parencites{CW2014}{Zbokov2018}, and functional penalisations, for instance, the MDL criterion has also shown its empirical usefulness \parencite{RPHR}. These motivate the formation of adaptive learning proposed here. 

Standard mathematical concepts and statistical notations are used in this paper. The concept of functional sets, as used in \textcite{Vapnik}, is highlighted later in \autoref{S3} as will be used frequently. Standard time series notations are used throughout, with the main reference on stationarity issues being \textcites{DF2}{TS2} and other ARIMA modellings being \textcites{H2}{TS3}. Bayesian hypothesis test is adapted at the level of \textcite{KPT2007}, with standard frequentist test being assumed at the level of \textcite{CB}.

 %\textcite{py} paper for model selection, with comments from BoE last year.

Details of the data cleaning, feature generation, and their exploratory analysis are written in \autoref{S2}. Modelling and learning proposals, together with the results are presented in \autoref{S3}, followed by an application to hypothesis testing in \autoref{S4} and the discussions in \autoref{S5}.

In terms of the key contribution from this paper, we engage with the existing methods on feature generation and traditional ARIMA formation, then propose a forecast-centric learning model, the adaptive learning, to approach model selections and post-estimation penalisation. Such a learning model helps to confirm the reachability of certain level of accuracy of the forecasting, improves the forecasting in volatile and non-stationary markets, and can also be applied into further analogies such as on its formation and hypothesis testing.
%\pagebreak
\section{Data}\label{S2}

\subsection{Data description, data cleaning  and computing deployment}
\begin{comment}
\begin{enumerate}
	\item Machines utilised and brief notes on parallel computing 
	\item Data cleaning and reasoning
	\item Feature generation
	\item Rolling-ADF tests on variables. 
	\item Extra example (with elaboration) possible but optional.
\end{enumerate}

\end{comment}

\subsubsection{Description, time brackets and VWM}

We base the statistical modelling on the intra-day price data of the CSI 300 Index Futures (hereafter called "the asset") provided by CIFCO Guangzhou. We focus the time range from 10th November 2017 to 17th April 2018, and there are 105 trading days in the range.

In a usual trading day, there are two trading sessions: one in the morning (0930---1130) and the other in the afternoon (1300---1500).\footnote{Further details can be retrieved from the exchange website: \url{http://www.cffex.com.cn/en_new/CSI300IndexFutures.html}.} In the dataset, we expect one or two raw entries within each second, while some omissions occur throughout. In each of the raw entry, the best bid and ask data together with the latest traded price and volumes are observed.
 As a decision to summarise the data, we divide each of the session into 24 brackets of 5-minute slots and index them by the end time, e.g. 0935 refers to the bracket from the first second of 0930 to the last second of 0934.\footnote{In HHMMSS format, that is from 093001 to 093500.} This is supported by the fact that noisiness and emptiness of the data and lack of transactions do exist. If summaries were to be made on a minute-level basis, while any brackets larger than 5 minutes would be less regarded as a high-frequency time series, as each session only has 2 hours.
 
 A summary of statistics and a histogram of the number of observations within each bracket are available in \autoref{Tab2.1} and \autoref{fig:B1} respectively in the appendix. \autoref{tab2} below serves as an example of translation between the brackets, thereafter "observations", the time, and in the financial environment.

\begin{table}[H]
	\centering
	\begin{tabular}{|c|c|c|c|}
		\hline 
		Number of Brackets & 12 & 24 & 48 \\ 
		\hline 
		Time & 1 hour & 2 hours & 4 hours \\ 
		\hline 
		Financial Remark & half trading session & 1 trading session & 1 trading day \\ 
		\hline 
	\end{tabular} 
	
	\caption{A guide between the number of observations, time, and financial meanings.}
	\label{tab2}
\end{table}

Within each bracket, we compute the Volume-Weighted-Mean (VWM) of the asset price. This is achieved by obtaining the arithmetic sum of the product of the trading volume and price in each of the raw entry, divided by the total volume.
Summary of statistics is supplied in \autoref{Tab2.1} in the appendix, and a line plot of the VWM is supplied in 	\autoref{fig:2-1} below.

\begin{figure}[h]
	\centering
	\includegraphics[width=\linewidth]{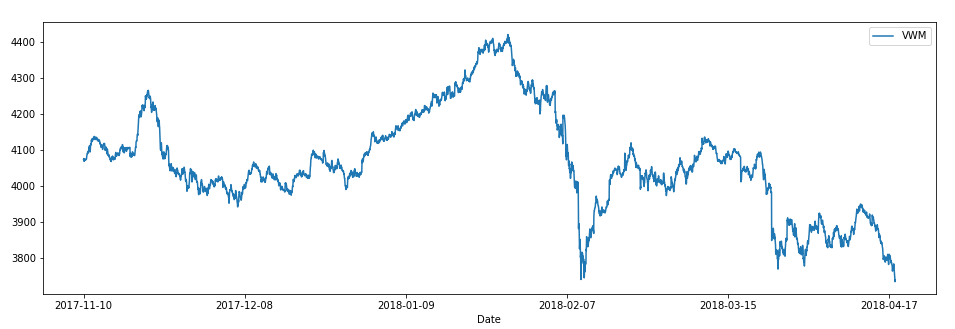}
	\caption{Plot of the VWM of the CSI 300 Index}
	\label{fig:2-1}
\end{figure}

\pagebreak
\subsubsection{Computing deployment}
Due to the computing complexity, cloud resources with parallel computing techniques are utilised. In particular, we deploy multi-core parallel computing tasks using the machines on Google Colab and AWS, which is achieved by centralising the function and distributing the parameters we wish to compute over different CPU cores, followed by result collections individually. Codes for execution and visualisation are written in Python 3.
%Further ... Appendix ...

\subsection{Feature generation from the order book}

As suggested by the literature review, we generate the Order Imbalance (OIB) and the Order Flow Imbalance (OFI) for each of the raw entries as follows:
\begin{subequations}
	\begin{align}
e_s^{OIB}  =& \frac{BQ_s-AQ_s}{BQ_s+AQ_s} \\
e_s^{OFI} = &  BQ_s \mathds{1}[BP_s \geq BP_{s-1}] - BQ_{s-1} \mathds{1}[BP_s \leq BP_{s-1}] \nonumber \\ &+ AQ_{s-1} \mathds{1}[AP_s\leq AP_{s-1}]- AQ_s \mathds{1}[AP_s \geq AP_{s-1}]
\end{align}
\end{subequations}
where we use $s$ as the index label for the time of the raw entry, $BQ_s, AQ_s$ as the best bid and ask quantities respectively, and $BP_s, AP_s$ as the best bid and ask prices respectively.

We provide a general interpretation without going deep into the theory here. For the OIB, when the bid quantity is relatively high, the OIB is more positive and vice versa if the ask quantity is relatively high. For the OFI, it can be seen as a signed contribution of the order book events to the supply or demand of the market of the asset. Say if someone buys passively through the current bid price, then $e_s=BQ_s-BQ_{s-1}$ represents the size of that order cancellation. If the bid price were to change --- depending on up or down, $e_s$ can represent the size of a price-improving order ($e_s=BQ_s$ if $BP_s > BP_{s-1}$), serving as a quantity for a rise in the demand; or the last order in the queue that was removed ($e_s=-BQ_{s-1}$ if $BP_s < BP_{s-1}$), thus a quantity for a drop in the demand. Likewise for the ask side symmetrically, where an increase in $AP$, for example, signifies a decrease in supply of the asset.

\begin{figure}[h]
	\centering
	\includegraphics[width=\linewidth]{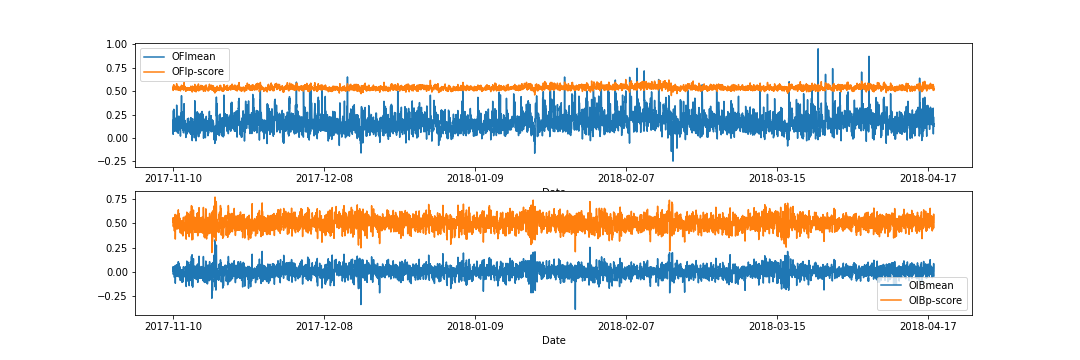}
	\caption{Plot of the four features generated}
	\label{fig:featuresline}
\end{figure}

Within each time bracket, we need to find representable summaries of statistics to represent the behaviour of each of the two features within, technically, $\{e_s^{OIB}, e_s^{OFI} | t-1 < t(s) \leq t \}$   where $t(s)$ indicates the time bracket that entry $s$ belongs to. Now, the mean within each bracket are the usual choice and is consistent with the intuition. In addition, we consider a p-score defined as \begin{equation}
p\text{\textendash}score_t:= \Phi \left(\frac{mean(\{e_s | t-1 < t(s) \leq t \})}{sd(\{e_s | t-1 < t(s) \leq t \})} \right)
\end{equation}
where $\Phi(\cdot)$ is the normal CDF and  $mean(\cdot)$ and $sd(\cdot)$ are the mean and standard deviation of the sequence. Summary of statistics of all of these feature generated are presented in \autoref{Tab2.1}, with a line plot below in \autoref{fig:featuresline}. The benefit of having a normal transformation, as seen from the plot or summary, is that the value can be restricted into a small range (theoretically $[0,1]$), which deals with any potential spiky moves of the fraction, while the cost is the decrease of variance associated with the increase of stability, which is not a huge trouble as it also brings time series models a benefit of stationarity.

\subsection{Exploratory data analysis}\label{EDA}
\subsubsection{Initial prices of the session: exemption and dummies}
Here we draw particular attention for the first and last observation of each session. Economically speaking, between the two trading sessions there could be large underlying events causing potential price movements, while the market is not open. This creates a high potential for the difference between the closing price of the previous session and the opening price of the current session to be large. We investigate these differences below. 

	\begin{table}[H]
		\centering
		\begin{tabular}{lccc}
			\toprule
			{} &     Day Gap &   Lunch Gap &          The rest \\
			\midrule
			count &   104 &     105 &  4830 \\
			mean  &     1.92 &      -0.44 &    -0.10 \\
			std   &    25.48 &       3.76 &     4.73 \\
			min   &  -134.03 &     -16.01 &   -52.51 \\
			max   &    79.36 &       6.75 &    32.49 \\
			\bottomrule
		\end{tabular}
		
		\caption{Summary of statistics for the gaps.}
		\label{Tab2.2}
	\end{table}

As shown in \autoref{Tab2.2}, the day gap, i.e. the difference between the first VWM observation in a morning's session and the last in the previous afternoon's session, is distributed much wider and have extreme values compared to the rest. This can be additionally supported by the	histogram in 
\autoref{fig:B1} in the appendix. While lunch gap, i.e. the gap between the start of the afternoon's session and the end of the morning's session, is small, to ensure consistency we exclude both of these gaps from estimation.
This is done by adding dummies when the time lands at these points.

%\pagebreak

To additionally ensure the stableness of forecasting models and that it has the ability to learn the new environment within each session before making forecasts, we exclude the first 6 observations, i.e. 30 minutes, of the session from forecasting. 

\subsubsection{Rolling ADF tests}

\begin{figure}[p]
	\centering
	\includegraphics[width=\linewidth]{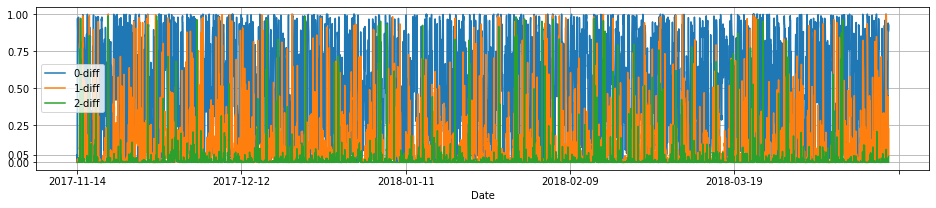}
	\includegraphics[width=\linewidth]{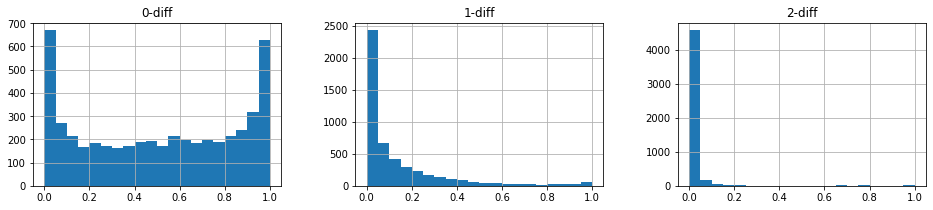}
	\includegraphics[width=\linewidth]{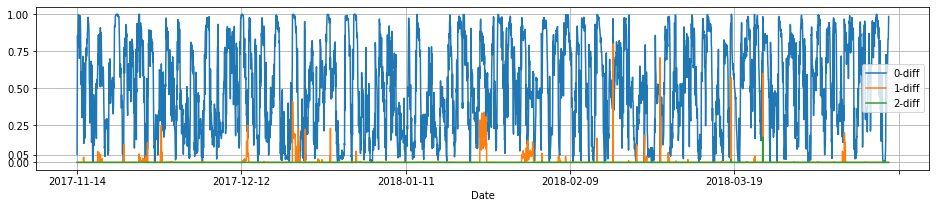}
	\includegraphics[width=\linewidth]{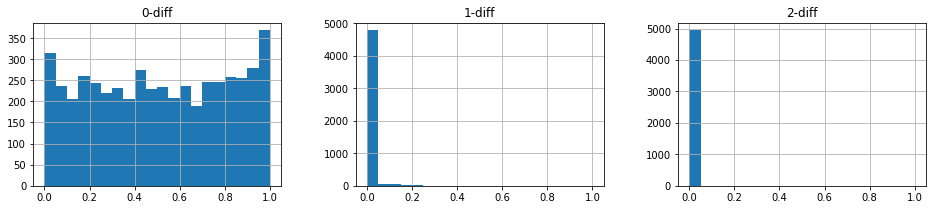}
	\includegraphics[width=\linewidth]{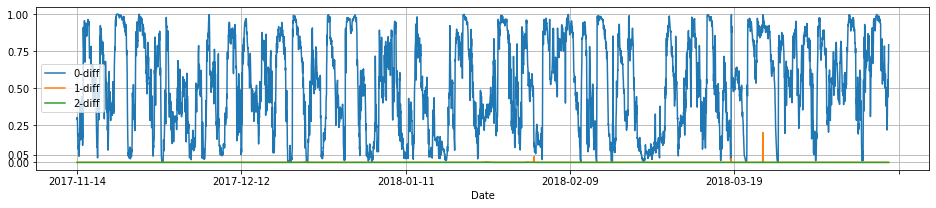}
	\includegraphics[width=\linewidth]{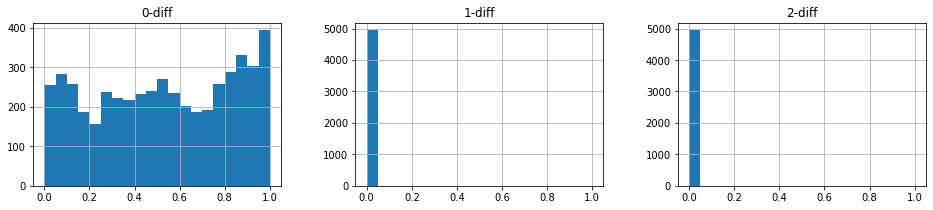}
	\caption{Results from rolling ADF tests for different window sizes: 12 (top two), 48 (centre two), and 96 (bottom two).}
	\label{ADFres}
\end{figure}

One crucial concern of time series is its stationarity. The approach undertaken to test the null hypothesis of unit root against the alternative hypothesis of stationarity, is via an Augmented Dickey-Fuller (ADF) test. Though, here we are interested in modelling the temporal relationships between variables, for which we are more concerned with the stationarity of the dependent variable, VWM of the price, in a short window. Hence we introduce the rolling ADF test, for which we collect the p-value, i.e. the probability of rejecting the null conditional on the null being true\footnote{Which is also the type I error.}, over time. 

Let $y_t$ be the VWM of the price at bracket $t$. Then, given a choice of window size $w$, we run an ADF test on the set $\{y_\tau \}_{\tau=t-w}^{t-1}$ for every $t$, and collect the result as $p_t(d;w)$ where $d$ is the level of difference.\footnote{Further details on the deployment of the test can be seen in \autoref{ADF}.} Interpretation of the result can be made by observing $p_t(d;w)$ against a critical value, which we take as 0.05 as usual.

Starting from $d=0$, if $p_t(0;w) < 0.05$, we conclude $\{y_\tau \}_{\tau=t-w}^{t-1}$ is stationary, else we seek for a higher order iteratively: until $p_t(d;w) < 0.05$ where we conclude  $\{\Delta^d y_\tau \}_{\tau=t-w+d}^{t-1}$ is stationary while $\{\Delta^{d-1}y_\tau \}_{\tau=t-w+d-1}^{t-1}$ is not. \footnote{$\Delta$ here is the difference operator, e.g. $\Delta y_t = y_t - y_{t-1}$ and $\Delta^ky_t = \Delta^{k-1}(y_t-y_{t-1}) \forall k \geq 2$.} 

We run this for three window sizes (12, 48, 96) and levels of difference (0,1,2) and draw, in \autoref{ADFres}, the line plot and histograms of the p-values for each of the combination.

The choppiness of the p-values for small-window sized data (when $w=12$) are significantly shown, while the larger ones seem stable with occasions where $p_t(1;w)>0.05$, meaning occasionally the 2nd level difference would be required for there to be a stationary model. 

These exploratory results help to decide how the time series model should be formed, as detailed in \autoref{S3}.
%\subsubsection{Rolling regressions}
\pagebreak

\subsubsection{The SR statistics: a trading perspective}\label{FinancialEDA}

A natural extension from a financial time series model is its profitability from trading. One good model should produce a reasonable return while maintaining suitable risks. This performance can be evaluated by the Sharpe Ratio (SR). Here we explain the construction towards such a statistics and provide baseline and feature-based results.

Let $P_t$ be the price of the asset at time $t$. Then the return for buying it at time $t$ and selling it at time $t+1$ is $\frac{P_{t+1}-P_t}{P_t}$. For each trading session after the forecasting exemption, we have 17 such opportunities, hence, given a theoretically zero-mean time series feature $\alpha_t$ for which the sign indicates the forecasted direction, we set the profit or loss in the trading session ($PL^{session}_s$) as 
\begin{equation}\label{eq4a}
PL^{session}_s := %\frac{1}{NoT_s} 
\sum_{t=t_s+1}^{t_s+17} sign(\alpha_t) \frac{P_{t+1}-P_t}{P_t}
\end{equation}
where $t_s+1$ locates the time index to the start of the session, and accordingly $t_s+18$ is the last observation of the session. For standard reporting on day profits or loss and further SR computation, we also produce the profit or loss in the trading day ($PL^{day}_d$) as \begin{equation}PL^{day}_d:=PL^{session}_{s_d} + PL^{session}_{s_d+1} \end{equation} where $s_d$ locates the morning session of a trading day $d$. %; and $NoT_s:=\sum_{t=t_s+1}^{t_s+23} \mathds{1}[\alpha_t \neq 0]$ is the number of trades one would make.
%Given our setting, we have 210 trading session in 105 days, hence for $PL^{day}_d$, we produce %its cumulative sum over time ($CPL_d$) and 
Now, the annualised $SR$ is defined as \begin{align}
%CPL_d=&\sum_{d^\prime=1}^{d} PL^{day}_{d^\prime} \\
SR=&\sqrt{252} \ \frac{ mean(PL^{day}_d)}{sd(PL^{day}_d)}
\end{align}
where $mean(PL^{day}_d)$ and $sd(PL^{day}_d)$ stand for the mean and standard deviation of $PL^{day}_d$ respectively.

In the baseline situation, we consider a buy-and-hold treatment, hence $PL^{session}_s$ is defined by simply buying from the start and selling at the last, thus $\frac{P_{t_s+18}-P_{t_s+1}}{P_{t_s+1}}$. Other statistics follows.

\begin{figure}[H]
	\centering
	\includegraphics[width=\linewidth]{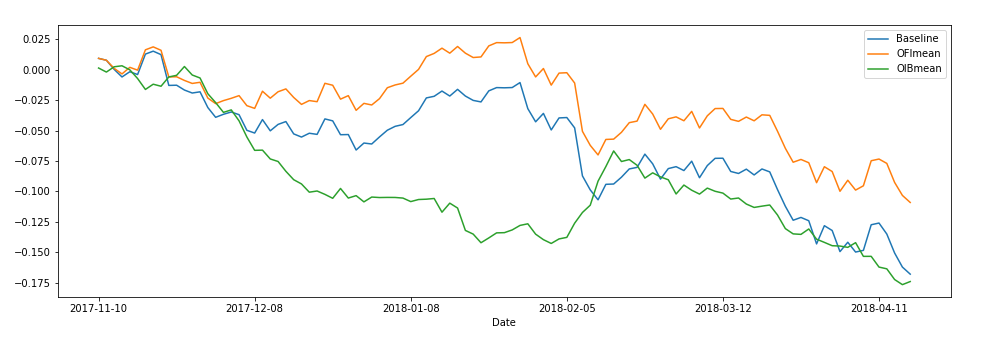}
	\caption{Plot of the cumulative $PL_d^{day}$ over time}
	\label{fig:2-2}
\end{figure}

The results of these are plotted in \autoref{fig:2-2}, which clearly shows the inability for the features themselves to achieve positive returns, while the baseline also performs badly.

\begin{table}[H]
	\centering
\begin{tabular}{lccc}
	\toprule
	{} &  Baseline &  OFI mean &  OIBmean \\
	\midrule
	mean ($PL_d^{day}$)  &   -0.0016 &  -0.0010 &  -0.0017 \\
	std ($PL_d^{day}$)  &    0.0093 &   0.0092 &   0.0063 \\
	min ($PL_d^{day}$)  &   -0.0394 &  -0.0397 &  -0.0184 \\
	max ($PL_d^{day}$)  &    0.0210 &   0.0206 &   0.0197 \\
	$SR$ & -2.71 & -1.79 & -4.20 \\

	\bottomrule
\end{tabular}
	\caption{Summary of statistics of $PL_d^{day}$ and $SR$}\label{tab3}
\end{table}

As a remark when SR is served as a performance metric later, the $\alpha_t$ is naturally set as $P_{t+1|t}-P_t$ where $P_{t+1|t}$ is the forecast of $P_{t+1}$ at time t, and \autoref{eq4a} can be interpreted as the trading profit or loss if one buys whenever the next price is forecasted to raise or sell otherwise.

%Statistics to supply: Cum Sum of return and Sharp Ratio.

\pagebreak
\section{Methodology and Results}\label{S3}
\subsection{General setting}
Let $y_t$ be the one-dimensional discrete time series of interest (the dependent variable), and let $\vx_t$ be the multi-dimensional discrete time series of features (explanatory variables). We are interested in forecasting the one-step-ahead future of the dependent variable conditional on the information up to time $t$, namely $y_{t+1|t}$.

In the common practice of time series, one studies the model of the underlying process  and then use the model to conduct forecasting (e.g. \textcite{PW2010}). Here we take a different approach: we first appreciate the conditional forecast as a value from a map that takes the information set ($\Phi_t$) and functional parameters ($\theta_t$,$h_t$), then build models to learn the appropriate parameters based on the previous observations. Mathematically, \begin{equation*}
y_{t+1 | t} = f(\Phi_t ; \theta_t ; h_t)
\end{equation*}
where $\Phi_t:=\{(y_t,\vx_t), (y_{t-1},\vx_{t-1}), ..., (y_{1},\vx_{1})\} = \{y_\tau,\vx_\tau | \tau\in [t] \}$, $\theta_t \in \Theta(h_t)$ is the parameter to be specified in the function, and $h_t \in H$ specifies the functional form, thus determines the parameter space $\Theta(h_t)$.\footnote{Here we note $\{f( \cdot ; \cdot ; h) | h\in H\}$ is a set of functions, this is the same notation as was used by \textcite{Vapnik}.} We manually design sensible models to construct $H$ and sensible learning methods on $\theta_t$ and $h_t$ to do good on reducing forecasting error --- we consider the usual MSE and MAE as the performance indicators.\footnote{See \autoref{MSE} for exact formulation.} MSE has a better theoretical foundation while MAE is more interpretable.\footnote{See the theoretical foundation of MSE, for instance, at the Corollary 8.17 of  \textcite{Klenke2013} where we view conditional expectation as projection.} We also consider the SR statistics %alongside other financial statistics introduced in 
%\autoref{FinancialEDA}
 which serve as an interpretable performance indicator in the context of financial time series.

At each of time $t$, $\vx_t$ is four dimensional: the first and second entries are, respectively, the mean of OIB and OFI; the third and fourth entries are, respectively, the p-score of OIB and OFI.

We index each $h\in H$  by $h(\iota, w, p,d,q)$, which controls an ARIMAX(p,d,q)-type of forecasting model with $w$ for window size and $\iota$  for the explanatory variables. For a given $p,d,q$, we consider a forecasting formula \begin{equation}\label{eq1}
y_{t+1|t}(h) = c_t+ g_{1,t}(p,d) y_t + g_{2,t}(q) \hat{\varepsilon_t}(h) + g_{3,t}(\vx_t)
\end{equation}
where $c_t$ is a constant, $g_1$ and $g_2$ are the appropriate ARIMAX operator functions: $g_1$ is specified by the autoregressive lag $p$ and difference parameter $d$, and $g_2$ is specified by the moving average lag $q$. $\hat{\varepsilon_t}(h)$ are the residuals from the model $h$. $ g_{3,t}(\vx_t)$ summarises the explanatory variables' contribution to forecasting.

In the followings, we first elaborate each of the specifications of $h$ with the associated method to pin down $\theta_t$, thus named "fixed models", then discuss adaptive learning models where $h_t$ can be time-varying by learning from the past. A general computing approach to obtain the result is shown in \autoref{alg1}.

\fbox{\begin{minipage}{\linewidth}
		\begin{algorithm}[H]
			\caption{Algorithm for obtaining the forecasts with a fixed $h$ (fixed models)}\label{alg1}
			\SetAlgoLined
			\KwIn{Data $\{\Phi_t\}_{t\in T}$ , specification of $h$, desired forecasting index set $T$, and validation data $\{y_{t+1}\}_{t\in T}$.}
			\KwOut{Forecasts $\{y_{t+1|t}(h)\}_{t \in T }$ and the performance metric.}
			\begin{enumerate}
				\item For $t\in T$, repeat:
				\begin{enumerate}
					\item Train parameters $\theta_t(h)$ on the windowed dataset $\Phi_t \setminus \Phi_{t-w}$, then obtain forecast $y_{t+1|t}(h)=\mE[y_{t+1}| \theta_t(h), \Phi_t,h]$
				\end{enumerate}
				\item Evaluate the performance metric.
		\end{enumerate}		\end{algorithm}
\end{minipage}}

\subsection{The fixed models: the univariate and the multivariate}

\subsubsection{Univariate framework}

In univariate models, the strategy to train the parameters $\theta^*_t \in \Theta(h)$ is rather classical: for a given $h(\iota, w, p,d,q)$ with with $\iota \in \{0,...,6\}$, we fit the following model in a $w$-windowed dataset:

\begin{minipage}{\linewidth}
\hspace{1cm} At time $\tau \in \{t-w+1,...,t\}$: 

\begin{equation}\label{eq2}
\Gamma_t(p) \Delta(d) y_\tau = \mu_t + \Phi_t(q) \varepsilon_\tau + <\beta_t(\iota), \vx_{\tau-1}> + dum_\tau  \  ; \ \varepsilon_\tau \sim iid N(0,\sigma_t^2)
\end{equation}

\end{minipage}

$\Gamma_t(p), \Delta(d), \Phi_t(q) $ are the lag operator functions under an ARIMAX (p,d,q) specification with constant $\mu_t$, e.g. $\Gamma_t(p)=(1-\gamma_{1,t}L - ... - \gamma_{p,t} L^p)$ where $L$ is the lag operator, i.e. $Ly_t=y_{t-1}$. The $dum_\tau$ term dynamically adds the number of required dummies as proposed in \autoref{EDA}.

The univariate model groups are specified by $\iota \in \{0,...,6\}$. In model group 0, we set $\beta(0):=(0,0,0,0)$, implying that the model is run in an ARIMA(p,d,q) fashion without explanatory variables. In model groups 1 and 2 we put $\beta(1):=(\beta_1,0,0,0)$, $\beta(2):=(0,\beta_2,0,0)$ meaning that we solely use the mean OIB in model group 1 and the mean OFI in model group 2. In model group 3, we utilise both the mean OIB and mean OFI, thus $\beta(3):=(\beta_1,\beta_2,0,0)$. Likewise for model groups 4 to 6 where in 4 and 5 we consider individually each of the p-scores, then in model group 6 we combine them. %Note that we do not consider linear + pscore interactions here, as otherwise the model has less interpretability.

We fit the model using a Maximum Likelihood Estimation (MLE) based on the specified dataset at each time $t$, thus obtain the relevant parameters to implement forecasting in \autoref{eq1}.

In the next paragraph we give an example to clarify the relationship between \autoref{eq1} and \autoref{eq2}. 

\subsubsection{Univariate example}

In this example we take $p=d=q=1, \iota=3$. Then \autoref{eq2} becomes
\begin{subequations}
\begin{equation}
(1-\gamma_t L) (1-L) y_\tau  =  \mu_t +  (1+\phi_tL) \varepsilon_\tau  + <\beta_t(3), \vx_{\tau -1}>+ dum_\tau    \  ; \ \varepsilon_\tau  \sim iid N(0,\sigma_t^2)
\end{equation}
and we may also write $<\beta_t(3), \vx_{t-1}>= \beta_{t,1} \vx_{t-1,1} + \beta_{t,2} \vx_{t-1,2} $ in scalar form. With these specifications, we can summarise all parameters\footnote{Apart from the dummies' term, which are straightforward to estimate.} to estimate as $\theta_t=(\mu_t, \gamma_t, \phi_t, \beta_{t,1}, \beta_{t,2},\sigma_t^2) \in \Theta(h)\subsetneq \mathbb{R}^6$, and in fact, by the standard time series set up we can pin down to the specified parameter region: $$\Theta(h)=\{\theta \in \mathbb{R}^6 | \theta_2 \in (-1,1),  \theta_6 \in (0,+\infty)  \}$$
Further into forecasting: once we obtained the appropriate $\theta_t^*\in \Theta(h)$, we proceed to \autoref{eq1}, which becomes
\begin{equation}
y_{t+1|t}=\mu_t^* + (1+\gamma_t^*)y_{t} -\gamma_t^* y_{t-1} + \phi_t^* \hat{\varepsilon_{t}} + <\beta_t(3)^*, \vx_t> 
\end{equation}
Equivalently, we can write $c_t=\mu_t^*$,  $g_{1,t}(p,d)=1+\gamma_t^*-\gamma_t^*L$, $g_{2,t}(q)=\phi_t^*$ and $g_{3,t}(\vx_t)= <\beta_t(3)^*, \vx_t> $.
\end{subequations}

\subsubsection{Univariate choices of parameter}
So far we explained the structure and strategy to train $\theta_t^*\in \Theta(h)$. Here we specify the choices of the model parameters. For the ARIMAX parameters, we put $p,q\in \{0,1,2\}$ and $d\in\{1,2\}$, with the choice of window sizes $w\in \{12,24,48,96\}$. We therefore have 72 models for each one of the seven univariate model groups, hence 504 models in total.

The reason for the window choices are from their corresponding financial meanings --- as one may note from the initial data cleaning (\autoref{tab2} in particular), 12 observations refer to one trading hour while 24 refers to a session. Likewise for 48, 96 which means one and two trading days respectively. As a result, $p,q$ may vary but rather restrictively due to the degrees of freedom, especially for smaller window sizes, hence the choice. The choice of $d$ can be both motivated from the literature and the rolling-ADF observations done previously (\autoref{ADFres}). While $d=1$ may be sufficient, in many occasions we need $d>1$ for  stationarity purposes, hence the choices for two potential values of $d$.

\subsubsection{Multivariate framework}
In multivariate model groups, we aim for the same forecasting formula as \autoref{eq1} but implement a vector training strategy: for a given $h(\iota,w,p,d,q)$ with $\iota \in \{7,...,12\}$, we train $\theta^*_t \in \Theta(h)$ by a $VARMA(p,q)$ on a stacked vector $S_t:=(\Delta(d)y_t, \vx_t) \in\mathbb{R}^5$. We fit the following model in a $w$-windowed dataset:

\begin{minipage}{\linewidth}
	\hspace{1cm} At time $\tau \in \{t-w+1,...,t\}$: 
	
	\begin{equation}\label{eq3}
	\Gamma_t(p) M(\iota) S_\tau = \bmu_t + \Phi_t(q) \bep_\tau + dum_\tau  \  ; \ \bep_\tau \sim iid N(0,\Sigma_t)
	\end{equation}
\end{minipage}

Here we first note the role of $M(\iota)$: it transforms the stacked vector $S_t$ to another which we subsequently perform VARMA on. In particular, $M(\iota)\in \{0,1\}^{n\times5}$ where $n\in\{2,3\}$ is the number of parameters we plan to have. Accordingly, 
$\Gamma_t(p)$ and $\Phi_t(q)$ are the lag operator functions under a VARMA (p,q) specification with $n$ dimensional variable, and the $dum_\tau$ term dynamically adds the number of required dummies as proposed in \autoref{EDA}.

The specification on $M(\iota)$ serves in the same spirit as was the $\beta_t(\iota)$ in \autoref{eq2}: it selects the relevant entries of $\vx_{t}$ to interact with $y_t$ and eventually contribute to the $g_{3,t}$ part of forecasting. 
For model groups 7 and 8, mean OIB and mean OFI, respectively, are the sole interaction being investigated, that is, $M(7), M(8)\in \{0,1\}^{2\times 5}$ and $M(7)_{1,1}=M(7)_{2,2}=1$ with the remaining entries being zero, $M(8)_{1,1}=M(8)_{2,3}=1$ with the rest being zero.

Similarly, training is done by MLE, and we proceed into an example.
\subsubsection{Multivariate example}
Consider $p=q=d=1, \iota=7$. Write $\widetilde{S_\tau}:=M(7)S_\tau=( (1-L)y_\tau , \vx_{\tau,1}) \in \mathbb{R}^2$. Then \autoref{eq3} becomes\footnote{Ignoring the dummy variables.}
\begin{subequations}
	\begin{equation}
	(I_2-\gamma_t L) \widetilde{S_\tau}  = \bmu_t +  (I_2+\phi_tL)  \bep_\tau    \  ; \  \bep_\tau  \sim iid N(0,\Sigma_t)
	\end{equation}
We note here $\gamma_t, \phi_t, \Sigma_t \in \mathbb{R}^{2\times 2} \cong \mathbb{R}^{4} $  therefore
$\theta_t=( \bmu_t, \gamma_t, \phi_t, \Sigma_t) \in \Theta(h)\subsetneq \mathbb{R}^{2} \times \mathbb{R}^{2\times 2\times 3} \cong \mathbb{R}^{14}$

Upon obtaining the appropriate $\theta_t^*\in \Theta(h)$, forecasting proceeds:
	\begin{equation}
	y_{t+1|t}=\bmu_{t,1}^* + y_t + \gamma_{t,1,1}^* (y_t-y_{t-1})  + \gamma_{t,1,2}^* \vx_{t,1} + \phi_{t,1,1}^*\hat{\bep}_{t,1}   + \phi_{t,1,2}^*\hat{\bep}_{t,2}
	\end{equation}
	
This is \autoref{eq1} with the specifications $c_t=\bmu_{t,1}^* $, 
$g_{1,t}(p,d)=1+\gamma_{t,1,1}^*(1-L)$, $g_{2,t}(q)= <\phi_t^* , (1,0)>$ and $g_{3,t}(\vx_t)=\gamma_{t,1,2}^* \vx_{t,1} $.
\end{subequations}

\subsubsection{Multivariate choices of parameter}
For the VARMA parameters, we put $p=1, \ q\in \{0,1\}, d\in \{1,2\}$ with $w\in \{48,96\}$. Hence 48 models are constructed in total. One may recognise this as a more restricted choice of parameters --- the choices of $w$ are limited to the larger ones due to the degrees of freedom. Take the previous example where the parameters to estimate is equivalent to 14 dimensional, it is not realistic to be implemented when window sizes are small. For the same reason, we cap $p,q\leq 1$ while if $p=0$ we get little meaning in the vector models, hence $p$ is fixed at 1 and $q$ may take one of the two values.

\subsection{Results from the fixed models}
\begin{figure}[H]
	\centering
	\includegraphics[width=\linewidth]{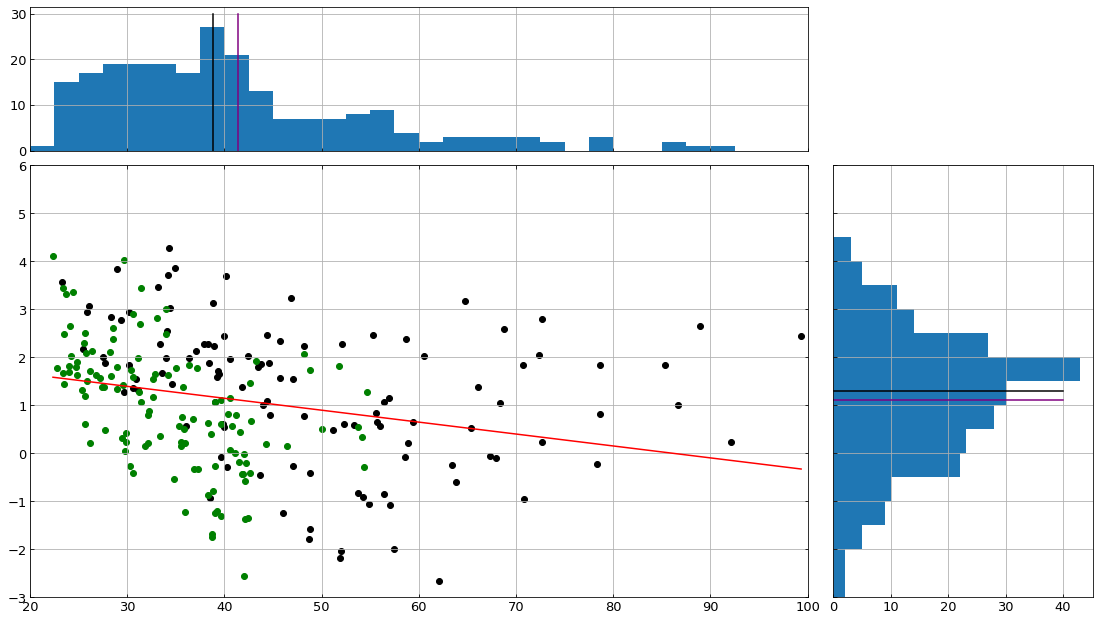}
	\includegraphics[width=\linewidth]{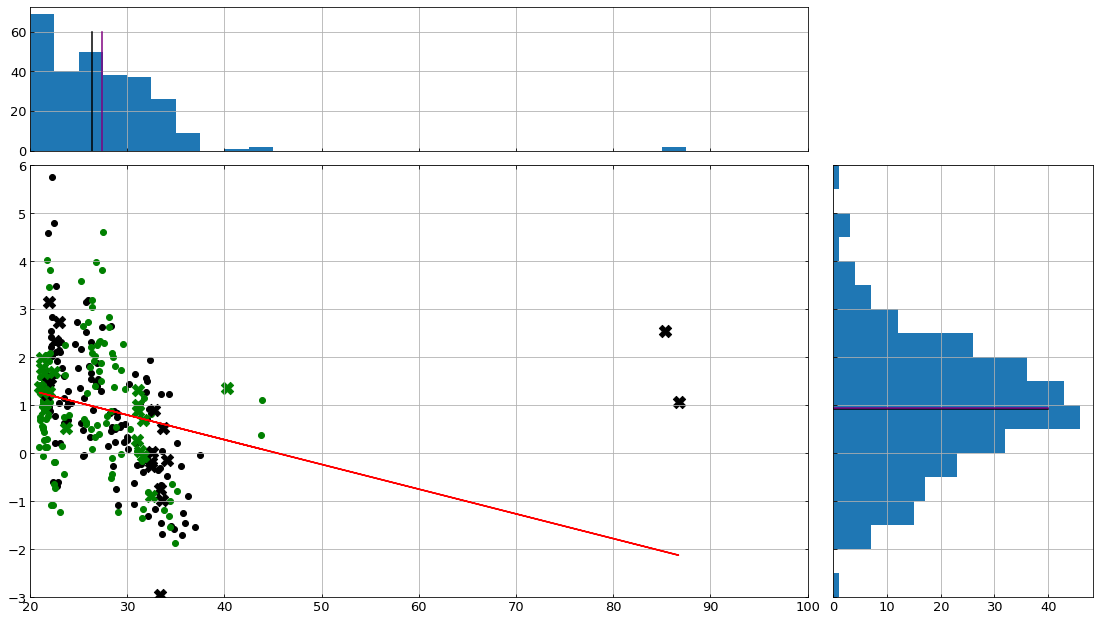}
	\caption{Scatter plots and histograms for each of the individual models. SR is on the vertical axis and the MSE is on the horizontal axis. The red line is the linear fit between these two. The window-size 12 results are plotted in black with the window-size 24 results plotted in green on the top panel, and likewise the window-size 48 in black and 96 in green on the bottom panel. Dots represent univariate models and crosses represent multivariate models. Black and purple lines in the histogram refer to the median and mean respectively.}
	\label{fig:B11}
\end{figure}

\begin{figure}[h]
	\centering
	\includegraphics[width=\linewidth]{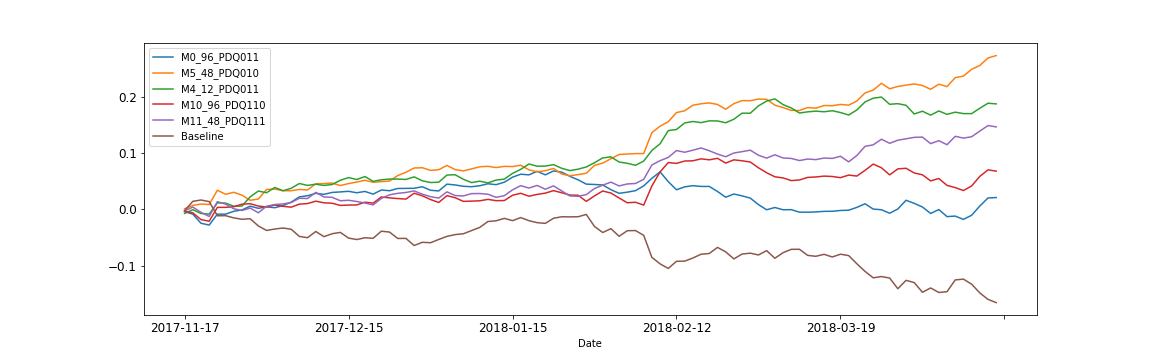}
	\caption{Plot of the cumulative $PL_d^{day}$ over time (Note: the legend is coded by model group number, window size, and $p,d,q$ values.)}
	\label{fig:pl}
\end{figure}

As a summary of the results thus far, we first plot the scatter and histograms in \autoref{fig:B11}, then, in \autoref{tab4}, \autoref{tab5}, and \autoref{fig:pl}, we produce tables and plots for the top-performing models under the MSE ranking and the SR. \autoref{TabB.1} in the appendix is also produced to summarise the relationship between model groups and explanatory variable(s). 

For a general result, we make scatter plots and histograms for all but the outliers models --- those which have an MSE greater than 100 are excluded from the plot. As can be observed from \autoref{fig:B11}, large-window models, in general, produce lower MSE, potentially benefited from its overall stability, while outstanding small-window may also have small MSE with large SR. The linear relationship between MSE and SR is weakly negative and with many points far below or above the fitted line. This supports the discrepancy as observed later, that some models may only perform well in one of the two metrics.

\begin{table}[H]
	\centering
	\caption{Univariate models: top 3 models ranked by MSE (upper) and SR (lower)}\label{tab4}
	\begin{tabular}{|c|c|c||c||c||c||}
		\hline 
		Model Group & $(p,d,q)$ & $w$ & MSE & MAE & SR \\ 
		\hline 
		0 & (0,1,1) & 96 & 20.97 & 3.34 & 0.13 \\ 
		\hline 
		0 & (1,1,0) & 96 & 21.03 & 3.35 & 0.69 \\ 
		\hline 
		5 & (0,1,1) & 96 & 21.22 & 3.36 & 0.99 \\ 
		\hline 
	\end{tabular} 
	\vspace{0.5cm}
	
	\begin{tabular}{|c|c|c||c||c||c||}
		\hline 
		Model Group & $(p,d,q)$ & $w$ & MSE & MAE & SR \\ 
		\hline 
		5 & (0,1,0) & 48 & 22.30 & 3.47 & 5.76 \\ 
		\hline 
		2 & (0,1,0) & 48 & 22.46 & 3.48 & 4.81 \\ 
		\hline 
		5 & (0,1,2) & 96 & 27.53 & 3.82 & 4.60 \\ 
		\hline 
	\end{tabular} 
	
\end{table}

For top-performing models, as seen from \autoref{tab4}, depending on which metric we use, the "top-performing" models could vary --- while models without any features (the top 2 of the upper table) perform well in MSE or MAE, their SR is rather low; with a slightly worse MSE and MAE models with features, here, in particular, the ones with either OFI mean or OFI p-score can obtain high SR, as seen from the lower table. 

An interesting observation about window size may also be made --- all of the models listed above are of size equal or greater than 48, similar out-performance may also be observed from the histograms of MSE in \autoref{fig:B11}. This corroborates with the classical statistical concern on stability, as the ones with smaller window sizes may have unstable estimations which occasionally induces large errors, therefore perform badly in MSE or MAE, but not necessarily in SR. 

We take a particular notice on one 12-window-sized model with $(p,d,q)=(0,1,1)$ from the model group 4 --- it has the fifth-highest ranking in the SR with MSE, MAE, and SR reported as 34.22, 4.22, and 4.28 respectively. This from another viewpoint shows the importance of having another performance metric --- while small-window models obtain drastic forecasts from time to time, their overall ability to forecast, or at least the direction (as the SR statistic is constructed in a way that it depends on the sign of the forecast rather on the magnitude of the forecast) may still be good. In fact, when looking at vector models below, we note this phenomenon to be rather significant as shown at the top row of the lower table of \autoref{tab5}. Indeed, smaller window sizes\footnote{Here we note that $w=48$ is relatively small in the context of vector models, due to the dimension of parameters it needs to estimate. One may observe from \autoref{fig:B11} that indeed the distribution of vector models for $w=48$ is much wider compared to the same windowed univariate models, and several extreme points exist.} cause instability, to an extent that 2 outliers of the 3600 \footnote{As checked in details of their distribution.} forecasts contribute largely to the bad-performing MSE and MAE. %\autoref{fig:B2} %to supply}

\begin{table}[H]
	\centering
	\caption{Multivariate models: the top 3 models ranked by MSE (upper) and SR (lower)}\label{tab5}
	\begin{tabular}{|c|c|c||c||c||c||}
		\hline 
		Model Group & $(p,d,q)$ & $w$ & MSE & MAE & SR \\ 
		\hline 
		10 & (1,1,0) & 96 & 21.03 & 3.37 & 1.39 \\ 
		\hline 
		11 & (1,1,0) & 96 & 21.24 & 3.37 & 1.78 \\ 
		\hline 
		8 & (1,1,0) & 96 & 21.26 & 3.38 & 1.99 \\ 
		\hline 
	\end{tabular} 
	\vspace{0.5cm}
	
	\begin{tabular}{|c|c|c||c||c||c||}
		\hline 
		Model Group & $(p,d,q)$ & $w$ & MSE & MAE & SR \\ 
		\hline 
		11 & (1,1,1) & 48 & 1603.69 & 4.42 & 3.24 \\ 
		\hline 
		8 & (1,1,0) & 48 & 21.95 & 3.43 & 3.15 \\ 
		\hline 
		8 & (1,1,1) & 48 & 22.97 & 3.51 & 2.73 \\ 
		\hline 
	\end{tabular} 
\end{table}

In terms of the top-performing vector models ranked by MSE, it is close to the univariate results with a slightly higher SR. While the top-performing vector models ranked by SR does not outperform the ones from univariate groups. This provides evidence that vector models do not perform outstandingly well in the context of one-step-ahead forecasting and windowed estimation. 

We also plot a cumulative PL of each of the top models in \autoref{fig:pl} along with the aforementioned one from the 12-window-sized and the baseline. As can be seen, the univariate models can perform better than the vector models, while the large-sized univariate model, i.e. the one with the lowest MSE, has a modest level of cumulative PL throughout. This could be explained by its ability to produce low errors, benefited from its stability from larger degrees of freedom, but not able to catch the time-varying change of the underlying parameters, hence the ability to forecast the sign and make trading profits is rather low.

\pagebreak
\subsection{Adaptive learning model groups}

\subsubsection{Principle of learning}
In the above twelve model groups, we fix $h_t$ constant throughout. Here, we consider an adaptive method to learn $h_t$. We also notice that, in the $h(11,48,1,1,1)$ for example, some drastic outliers could worsen the performance, especially if evaluated by the MSE. Hence we first shrink the functional set $H$ by assigning certain error and outlier handling ability as below: 
\begin{equation}\label{AL1}
\tilde{H_t}=\left\{h\in H \bigg| y_{t+1|t} (h) \text{ is positive and } \left| \frac{y_{t+1|t} (h)}{y_t} -1 \right| <0.05 \right\}
\end{equation}

This makes sure the set of functions we are selecting from are not outliers. Now, let time index $W$ be large enough so we can run all the previous model groups and obtain a suitable amount of  forecasting error $\epsilon_{\tau+1}(h)$ from each of the model $h$, which is defined as follows. At any $t\geq W$, we have access to models in the group 0 to 12, which produce forecasts $y_{\tau+1|\tau} (h)$
for each of the $h\in H, \ t-1 \geq \tau \geq W$ . We write $\epsilon_{\tau+1}(h):=y_{\tau+1}-y_{\tau+1|\tau} (h)$ and the adaptive learning aims to learn from the errors available up to time $t$, together with other information available, to make a decision on the model to employ at time $t$.

A general strategy to train $h_t$ at time $t$ is to construct a loss function $\ell (\Phi_t,H)=\ell (\Phi_t,h, H \setminus \{h\})$ and solve the appropriate minimisation problem: \begin{equation}\label{AL2}
h_t^* = \argmin_{h\in \tilde{H_t}} \ell (\Phi_t,h,H \setminus \{h\})
\end{equation}

The associated computing procedure is supplied in \autoref{alg2}.

\fbox{\begin{minipage}{\linewidth}
		\begin{algorithm}[H]
			\caption{Algorithm for obtaining the forecasts with a time-varying $h_t$}\label{alg2}
			\SetAlgoLined
			\KwIn{Data $\{\Phi_t\}_{t\in T}$, functional sets $H$, specification of the loss function $\ell (\cdot,\cdot)$, desired forecasting index set $T$, and validation data $\{y_{t+1}\}_{t\in T}$.}
			\KwOut{Forecasts $\{y_{t+1|t}(h_t^*)\}_{t \in T }$ with the associated functions $\{h_t^*\}_{t\in T}$, and the performance metric.}
			\begin{enumerate}
				\item For $t\in T$, repeat:
				\begin{enumerate}
					\item Produce $ \tilde{H_t}$ according to \autoref{AL1}.
					\item Evaluate and execute the minimisation given by \autoref{AL2}. Then get $h_t^*$ with $y_{t+1|t}(h_t^*)=\mE[y_{t+1}| \theta_t(h_t^*), \Phi_t,h_t^*]$.
				\end{enumerate}
				\item Evaluate the performance metric.
		\end{enumerate}		\end{algorithm}
\end{minipage}}

In the followings, we consider two groups of specifications of the loss function, with the first one motivated from \textcite{py}.

\subsubsection{Adaptive learning on errors}

 In model group 13, we construct \footnote{Now and thereafter, we abbreviate $\epsilon_{\tau}(h)$ as $\epsilon_{\tau}$ where $h$ is clearly emphasised on the left hand side of the equation.} \begin{equation}
\ell (\Phi_t,h,H \setminus \{h\})= L^{global}(\{\epsilon_{\tau}\}_{\tau=W}^t)= \sum_{t-47\leq \tau \leq t} \lambda ^{t-\tau}  \ L^{local}(|\epsilon_{\tau}|)
\end{equation}
The above line defines the loss by solely focusing on the errors in the past 48 observations, which are diminishing geometrically at a rate $\lambda\in(0,1]$. The local loss function $ L^{local}(\cdot)$ is specified as a continuously differentiable combination between zero, square loss, and absolute loss, with constants $C_1,C_2$ where $0\leq C_1 \leq C_2 \leq \infty $: \footnote{Note also that if $C_1=0$ and $C_2=\infty$, the local loss becomes proportional to the MSE contribution, similar to the ones proposed by \textcite{py}.}
\begin{equation}
L^{local}(x ; C_1,C_2)=
\begin{cases}
(C_2-C_1) x + \frac{C_1^2-C_2^2}{2} & \text{if $x>C_2$}\\
\frac{(x-C_1)^2}{2}  & \text{if $C_2\geq x >C_1$}\\
0 & \text{otherwise}
\end{cases}       
\end{equation}

Benefited from the high-frequency dataset, differently from \textcite{py}, we here place a time-varying constants at which the losses switch --- we trial different quantiles of $\{|\varepsilon_{t}(h)| \big| h\in H \}$ to set $C_1(t), C_2(t)$. E.g. $C_1(t)$ can be the 25\% quantile of $\{|\varepsilon_{t}(h)| \big| h\in H \}$ and $C_2(t)$ can be 50\% or 75\% quantile of $\{|\varepsilon_{t}(h)| \big| h\in H \}$. Note here that the quantiles have the benefit of outliers-resilient as it relates to the distribution rather than expectation of the set. The time-varying parameter here assists the local penalisation to be done in a time-varying manner and thus the minimisation process.

As to the selection of parameters, we consider $\lambda \in \{0.8,0.85,0.9,0.95,0.99,1\}$ and $25\% , 50\% ,75\%$ quantiles for the $C_1,C_2$. Some interpretation on the $\lambda$ can be made based on its logarithmetic and exponential properties --- the half period of 0.95 for example, is around 13.5 thus having $\lambda=0.95$ essentially reviews the past 14 observations with little role being played by the further ones, while $\lambda=0.8$ is more extreme, as its half period is only about 3.1. A general table is supplied in \autoref{TabB.2}.

\subsubsection{Adaptive learning with functional awards and penalties}

One concern about purely focusing on the forecasting error is the potential to misfit as the functional form plays a role in the degree of freedom and may  also be of importance when selecting which model to adapt as the most appropriate one for $h_t^*$. The classical approach that adds penalisation on the model selection criteria is from  \textcite{Akaike1974}, and later the Lasso methods, in particular, the fused lasso \parencite{Tibshirani2005}.  Though we are different from the previous methods as we focus on the out-of-sample loss rather than the in-sample likelihood. 

Here, in model group 14, we take a more time-varying approach: we consider a penalisation or reward, depending on the particular design, between the function of concern $h$ and the previous choice $h_{t-1}^*$, write as $D(h,h_{t-1}^*)$.

\begin{equation}\label{AL4}
\ell (\Phi_t,h,H \setminus \{h\})= L^{global}(\{\epsilon_{\tau}\}_{\tau=W}^t) + D(h,h_{t-1}^*)
\end{equation}

In the followings, we first consider, in type-1, purely penalising the difference in each of the variables that form $h$, i.e. the time series parameters and the window sizes.
\begin{subequations}
	
In type-1 we design \begin{equation}
D(h,h_{t-1}^*;C_3,C_4)= C_3|p+d+q-p^*_{t-1}-d^*_{t-1}-q^*_{t-1}|  + C_4 |w-w^*_{t-1}|
\end{equation}
where $C_3, C_4 > 0$.

As specified, we pool the $p,d,q$ together and, due to the size difference, treat $w$ separately. The fact we pool $p,d,q$ together can be appreciated as the change in the complexity, in particular, number of lags and differences in total. Another potential way is to separate them and penalise the change one by one.\footnote{For instance, $C_{3,1}|p-p^*_{t-1}|  + C_{3,2}|d-d^*_{t-1}|   + C_{3,3}|q-q^*_{t-1}|  $.} What is also interesting to consider is the fact that we may still have a preference to large-window sized models due to their stability, hence an additional reward can be made to encourage switches into smaller window-sized models and differenced orders, hence the followings for type-2 and 3:
 \begin{equation}
D(h,h_{t-1}^*;C_3,C_4,C_5)= C_3|p+q-p^*_{t-1}-q^*_{t-1}|  + C_4 (48-w) + C_5 |d-d^*_{t-1}|
\end{equation}
where $C_3 > 0 , \ \ \ C_4, C_5 <0$ in type 2. Now, in type-3 we put \begin{equation}
D(h,h_{t-1}^*;C_3,C_4,C_5)= C_3|p+q-p^*_{t-1}-q^*_{t-1}| \mathds{1}[d=d^*] + C_4 (48-w) + C_5 |d-d^*_{t-1}|
\end{equation}
where $C_3 > 0 , \ \ \ C_4, C_5 <0$.

Both of the specifications have a strictly increasing reward for smaller window sizes as small-window sized ones get a more negative value on the $ C_4 (48-w)$ term, which proceed into a beneficial status at the minimisation stage. The difference between the type-2 and 3 is that in type-3, penalisation on the $p,q$ terms are only applied if $d$ stayed the same --- this paves path for potential switches in $d$, which may be desirable when instability breaks out and later finished.
\end{subequations}

We inherit the parametrisation in model group 13, while for the $C_3,C_4,C_5$, we consider the following time-varying parametrisation. Let $$l_t^*:=\min_{h \in \tilde{H_t}} L^{global}(\{\epsilon_{\tau}\}_{\tau=W}^t)$$ and we set $C_3(t),C_4(t),C_5(t) \propto l_t^*$.
In particular, for type-1, we put $$C_3(t)= \frac{1}{10} l_t^*, \ C_4 = \frac{1}{168}l_t^*$$ and for type-2 and 3, $$C_3(t)= \frac{1}{8}l_t^*,  \ C_4(t)= - \frac{1}{72} l_t^*, \  C_5(t)=- \frac{1}{2}l_t^*$$

The rationale behind these particular fractions is that we aim to control the maximum loss being added from each term of the $D(h,h_{t-1}^*)$ to be half of $l_t^*$, and likewise for the magnitude of the rewards in the type-2 and 3.

\subsection{Results from adaptive learning model groups}

\begin{table}[h]
	\centering
	\caption{Top models from each model group-type, ranked by MSE (upper) and SR (lower)}\label{tab6}
	\begin{tabular}{|c|c|c|c||c||c||c||}
		\hline 
		Model Group & type & $\lambda$ &$(C_1,C_2)$& MSE & MAE & SR \\ 
		\hline 
		14& 1 &1 &(50\%,75\%)  &22.05&3.42& 1.54\\ 
		\hline
		13 &  & 1 & (50\%,75\%) &  22.14& 3.42  & 2.07 \\ 
		\hline
		14 &2 &0.99&(50\%,75\%) &22.32& 3.45& 1.15\\ 
		\hline 
		14 &3 &0.99 &(50\%,75\%) &22.38 &3.45 &1.37 \\ 		
		\hline

	\end{tabular} 
	\vspace{0.5cm}
	
		\begin{tabular}{|c|c|c|c||c||c||c||}
			\hline 
			Model Group & type & $\lambda$ &$(C_1,C_2)$& MSE & MAE & SR \\ 
		\hline 
14& 1 &0.85 &(50\%,75\%) &22.52&3.47&2.90 \\ 
\hline 

14 &3 & 0.8&(25\%,50\%) &33.14 & 4.15&2.63 \\ 		
			\hline 
			13 &  & 0.85 & (50\%,75\%) &  22.51& 3.47  & 2.56 \\ 
			\hline 
14 &2 &0.8&(25\%,50\%) &28.33& 3.85&2.41 \\ 
\hline 
		\end{tabular} 
	
\end{table}

For each of the type in model group 14 and throughout the model group 13, we select the best model ranked in either MSE or SR and list them above in \autoref{tab6}. Compared against \autoref{tab4}, we see the adaptively learnt models may achieve similar, though no better results, compared to the ones in the fixed models, both when measured in MSE and SR. This motivates the later subsection where we look closer into some of the periods that adaptively learnt models outperform the best one from the fixed model groups.

\begin{figure}[h]
	\centering
	\includegraphics[width=1\linewidth]{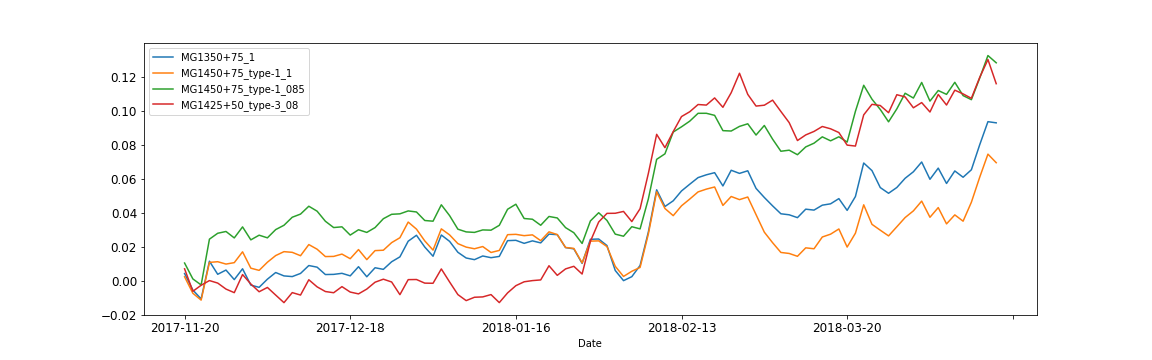}
	\includegraphics[width=1\linewidth]{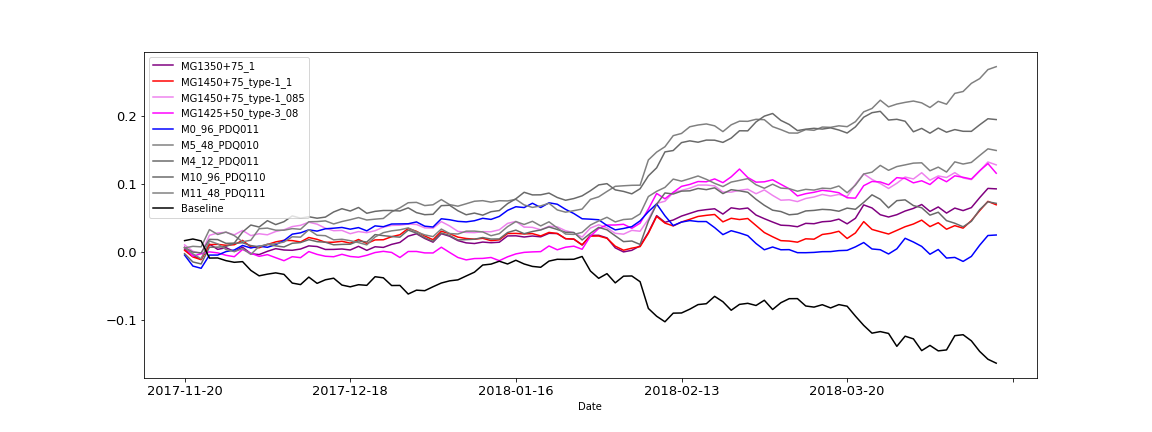}
	\caption{Plot of the cumulative $PL_d^{day}$ over time}
	\label{fig:pl12}
\end{figure}

To have a more financial comparison, we observe in \autoref{fig:pl12}, the cumulative PL plotted first within themselves and then with the top ones from the fixed model groups. Compared to the best model measured in MSE, which performs poorly in the SR, the adaptively learnt models obtain higher cumulative returns while not high enough compared to the ones which are the best individual models measured in SR.

\begin{table}[H]
	\centering
	\caption{Summary of statistics of $PL_d^{day}$ (columns 1 to 4, within which all entries are multiplied by 100 for cleanness) and $SR$ (column 5)}\label{tab7}
	\begin{tabular}{l|c|c|c|c||c}
		{Model code} &   $ 100 \times$ mean  & $ 100 \times$    std & $ 100 \times$     min & $ 100 \times$    max &      $SR$ \\
		\hline 
		M5\_48\_PDQ010         &  0.28 & 0.74 & -1.02 & 3.72 &  5.96 \\
				\hline 
		M4\_12\_PDQ011         &  0.20 & 0.71 & -1.55 & 2.34 &  4.38 \\
				\hline 
		M11\_48\_PDQ111        &  0.15 & 0.73 & -1.17 & 2.63 &  3.29 \\
			\hline 	MG14\_50+75\_type-1\_085 &  0.13 & 0.71 & -1.10 & 2.70 &  2.89 \\
		\hline 		MG14\_25+50\_type-3\_08  &  0.12 & 0.71 & -1.42 & 2.30 &  2.61 \\
			\hline 	MG13\_50+75\_1          &  0.09 & 0.73 & -1.47 & 2.48 &  2.05 \\
			\hline 	MG14\_50+75\_type-1\_1   &  0.07 & 0.73 & -1.16 & 2.48 &  1.53 \\
			\hline 	M10\_96\_PDQ110        &  0.07 & 0.79 & -1.26 & 3.42 &  1.46 \\
			\hline 	M0\_96\_PDQ011         &  0.03 & 0.70 & -1.74 & 1.95 &  0.58 \\
		Baseline             & -0.17 & 0.95 & -3.94 & 2.10 & -2.76 \\
		\bottomrule
		
	\end{tabular}
	\vspace{0.3cm}
	\begin{flushleft}Note about the number of days: after some data for the initialisation of models, we have 99 trading days starting from 20th November 2017, which explains the small difference between the baseline data here and the ones in \autoref{tab3}.
		\\
		Note about the model code: for fixed models, they are coded by model group, window size, and $p,d,q$ parameters; for the learning models, they are coded by model group, $C_1,C_2$ in percentage, type number (if applies), and $\lambda$.
	\end{flushleft}
	
\end{table}

Another financial comparison can be made from \autoref{tab7}, that the standard deviation, min, and max of the daily return from the adaptive learning models are all in line with the top-performing fixed models, which intuitively explains the stability of these adaptively learnt models.

\begin{figure}[H]
	\centering
		\includegraphics[width=0.3\linewidth]{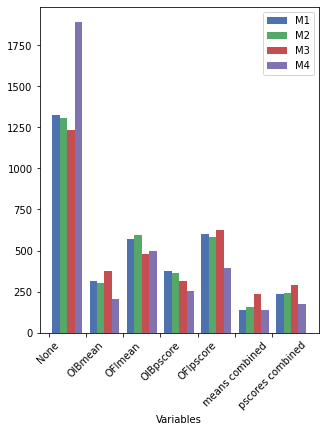}
	\includegraphics[width=0.6\linewidth]{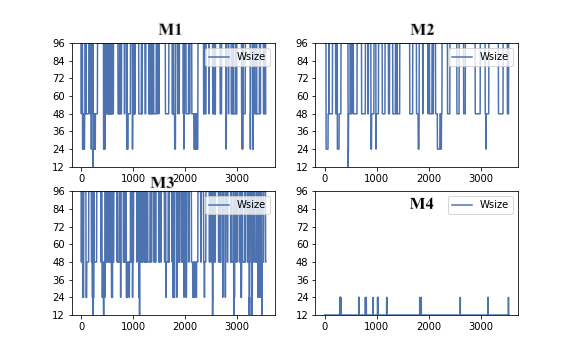}
	\includegraphics[width=0.25\linewidth]{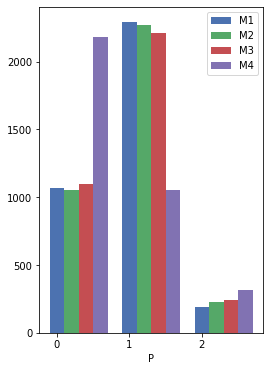}
	\includegraphics[width=0.25\linewidth]{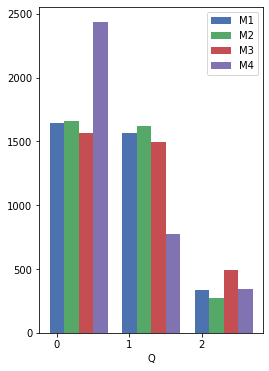}
		\includegraphics[width=0.2\linewidth]{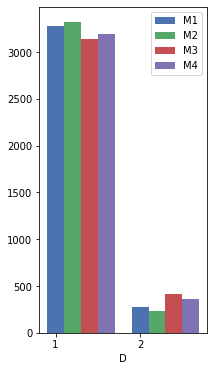}
				\includegraphics[width=0.2\linewidth]{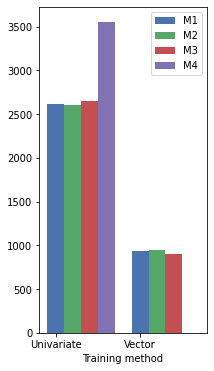}
	\caption{Plots of the formation of the four model groups: variables (top left), window sizes (top right), $p,d,q$ parameters and their training method (bottom)}
	\label{fig:hist4}
\end{figure}

The benefit of adaptive learning models is one could look into the formation of each of the models. As shown in \autoref{fig:hist4}, we may see the preferences of the model parameters through the period --- univariate models encapsulate a majority, $d=2$ is occasionally visited, and there is a good blend of usage of explanatory variables and the other time series parameters.

The choices of window sizes, depending on the design, may vary largely --- in particular, M4 has a strong preference towards the smaller ones and, in terms of explanatory variables, it prefers to use none of them.

In the following subsection, we take a closer look at periods when adaptive learning results outperform the fixed models.

\pagebreak

\begin{table}[H]
	\centering
	\caption{Model number assigned for \autoref{fig:hist4}}\label{tabhist4}
	\begin{tabular}{|c|c|c|c|c|}
		\hline 
Model number assigned for \autoref{fig:hist4}	&	Model Group & type & $\lambda$ &$(C_1,C_2)$ \\ \hline
M1	&	13 &  & 1 & (50\%,75\%)  \\ 
		\hline 
M2	&	14& 1 &1 &(50\%,75\%)  \\ 
		\hline
M3
&		14& 1 &0.85 &(50\%,75\%)  \\ 
		\hline 
	M4	
&		14 &3 & 0.8&(25\%,50\%) \\ 		
		\hline 
	\end{tabular} 
	
\end{table}

\pagebreak
%\pagebreak
\subsection{A closer look at adaptive learning results}
\begin{figure}[H]
	\vspace{-1cm}
	\centering
			\includegraphics[width=0.9\linewidth]{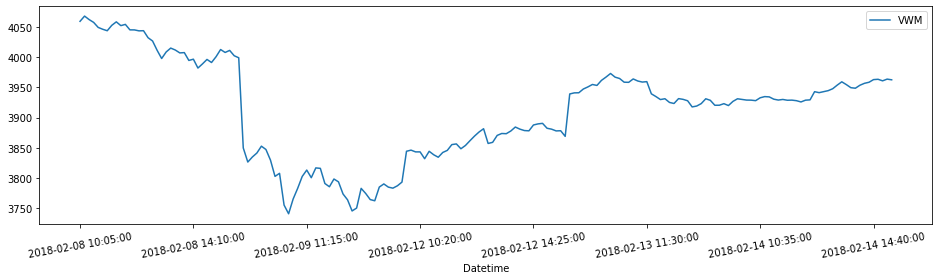}
		\includegraphics[width=0.9\linewidth]{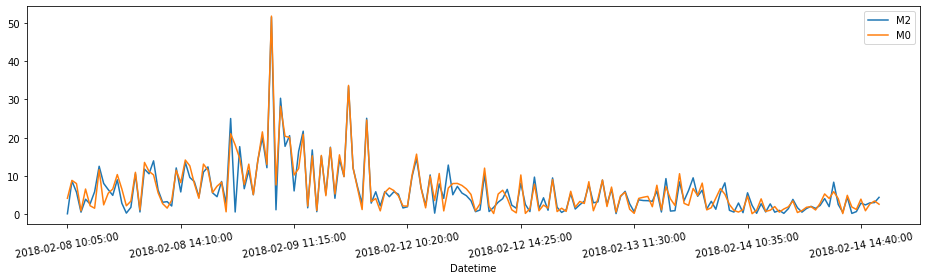}
	\includegraphics[width=0.9\linewidth]{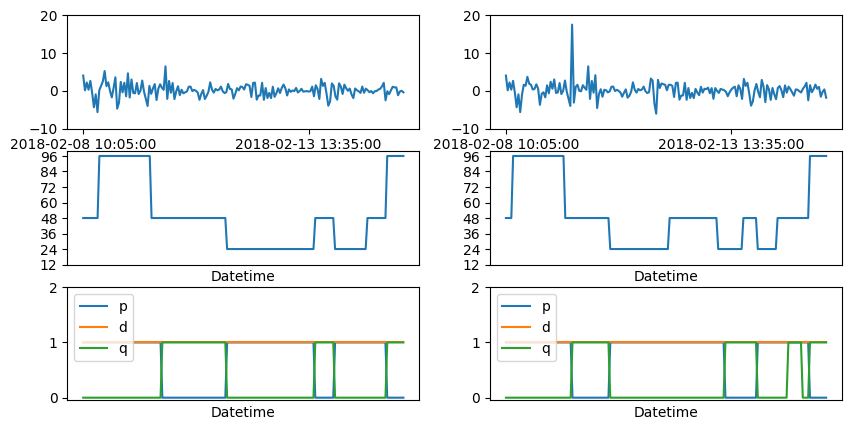}
	\includegraphics[width=0.3\linewidth]{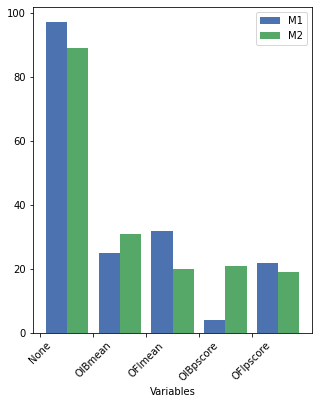}
	\includegraphics[width=0.2\linewidth]{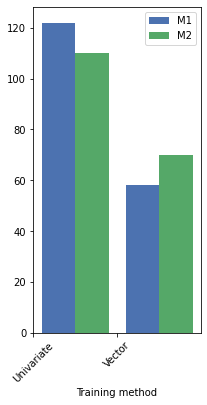}
	\caption{The first plot: the price during the 5 trading-day period; the second plot: plot of absolute forecasting errors of $M_0$ and $M_2$; the bottom four plots: error comparison and model formations of $M_1$ (left) and $M_2$ (right).}
	\label{fig:hist6}
\end{figure}

We first observe a period when prices are volatile and most of the models obtain large forecasting errors. We zoom into a five trading day period starting from 8th February 2018. As plotted in the top of \autoref{fig:hist6}, there is a large drop and subsequently big fluctuations around. We label $M_0$ as the best-performing individual model in terms of MSE, which is the sized 96, $(p,d,q)=(0,1,1)$ model with no explanatory variable, and as we see from the second plot of \autoref{fig:hist6}, the forecasting errors can be spiky and occasionally large, contributing a large MAE and MSE on average, as tabled in \autoref{tab9}.

We observe from two outstanding models from the model group 14 --- labelled as $M_1$, the type-3 with $\lambda=1$ and $(C_1,C_2)=(50\% , 75\%)$ and labelled as $M_2$, the type-2 with $\lambda=0.9$, $(C_1,C_2)=(50\% , 75\%)$ while $C_5(t)$ adjusted to $-l_t^*$.

\begin{table}[H]
	\centering
	\caption{Summary of the absolute forecasting errors (columns 1 to 3) and squared forecasting errors (columns 4 to 6)}\label{tab9}
\begin{tabular}{c|ccc||ccc}
	Model and error &  $M_0$ ($|\epsilon_t|$) &   $M_1$ ($|\epsilon_t|$)   &    $M_2$ ($|\epsilon_t|$) &  $M_0$ ($\epsilon_t^2$)  &  $M_1$($\epsilon_t^2$) &  $M_2$($\epsilon_t^2$)  \\
	\hline
mean  &     6.30 &   6.14 &   6.08 &          83.99 &       83.44 &       81.94 \\	\hline
min   &     0.05 &   0.03 &   0.03 &           0.00 &        0.00 &        0.00 \\	\hline
	max   &    51.81 &  51.52 &  51.52 &        2684.70 &     2654.32 &     2654.32 

\end{tabular}

\end{table}

As observed from \autoref{tab9}, the MAE and MSE in this period are generally high, while the adaptively learnt models have reduced them to a certain extent. The second plot of \autoref{fig:hist6} plots the absolute forecasting error of $M_0$ and $M_2$, and the third row of plots of \autoref{fig:hist6} show the level of improvement (if positive) or worsening (if negative) from $M_0$ to $M_1$ on the left, and from $M_0$ to $M_2$ on the right.

The selection of window sizes may also explain the source of improvement --- shorter ones are selected during a few periods when the prices are relatively unstable, and the relevant $p,q$ parameters are changed throughout. These details can be further seen from the bottom of \autoref{fig:hist6}. Among the $M_1$ and $M_2$, some difference can also be observed in the variable selection and training method, though most of the time no explanatory variable and univariate training methods are preferred.

The trend that adaptive learning models seem to perform better in the non-stationary part of the data motivates another review on the performance of adaptive learning models. Here we consider the non-stationary data as observed in the \autoref{EDA} --- there were time when, at the level of window-sized 12 that we reject the null hypothesis of the ADF tests at 5\% significance level for both 0-diff, 1-diff, and 2-diff, meaning that $y_t$ is not stationary even after twice the difference. There are 152 observations of this nature that intersect with the forecasting set, and below in \autoref{tab9-2}, we provide a summary of statistics for the forecasting errors.

\pagebreak
	\thispagestyle{empty}

\begin{table}[H]
	\centering
	\caption{Summary of the absolute forecasting errors (columns 1 to 2) and squared forecasting errors (columns 3 to 4)}\label{tab9-2}
	\begin{tabular}{c|cc||cc}
		Model and error &     $M_0$ ($|\epsilon_t|$) &     $M_1$ ($|\epsilon_t|$) &  $M_0$ ($\epsilon^2_t$) &  $M_1$ ($\epsilon^2_t$)  \\
		\midrule
		mean  &   3.30 &   3.39 &     26.28 &     24.73 \\
		min   &   0.01 &   0.05 &      0.00 &      0.00 \\
		max   &  24.69 &  22.86 &    609.76 &    522.41 \\
		\bottomrule
	\end{tabular}
\end{table}

\begin{figure}[H]
	\centering
	\includegraphics[width=0.9\linewidth]{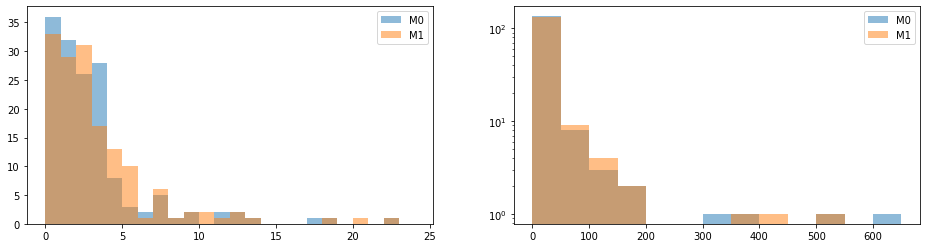}
	\caption{Histograms of the absolute errors (left) and the squared errors in $\log_{10}$ scale (right)}
	\label{fig:hist7}
\end{figure}

In \autoref{tab9-2}, we see the best-fixed model ($M_0$) to have moderate MAE and MSE while the adaptively learnt one ($M_1$) has slightly smaller MSE benefited from its avoidance from the large errors. The histogram in \autoref{fig:hist7} also supports such evidence --- errors from $M_1$ have less distribution on the right tail.

\begin{figure}[H]
	\centering
	\includegraphics[height=0.25\textheight]{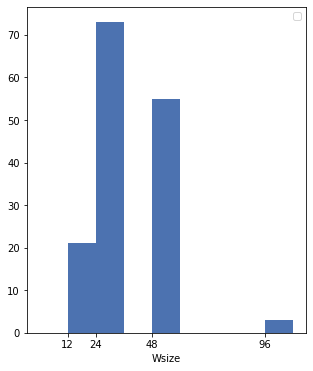}
		\includegraphics[height=0.25\textheight]{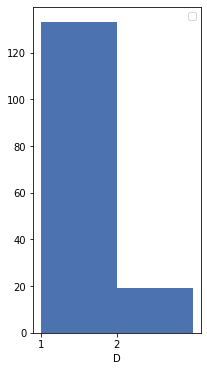}
			\includegraphics[height=0.25\textheight]{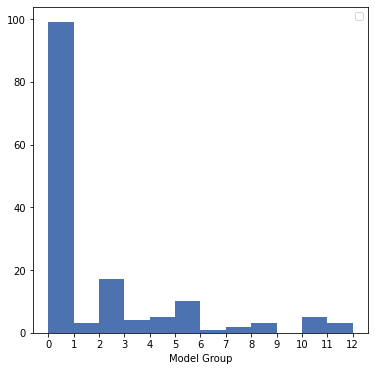}
	\caption{Histograms of the window sizes (left), difference order (centre), and model group (right)}
	\label{fig10}
\end{figure}

The formation of the adaptively learnt model presented here can be found below in \autoref{fig10}: it has a high proportion of model group 0, meaning there are a vast majority of models being purely ARIMA without explanatory variables, while the window sizes tend to be the smaller ones and avoids the 96. The difference order is usually taken at 1, similar to the others observed previously.

%\pagebreak
\section{Application to Hypothesis Testing on models}\label{S4}
Having had adaptively learnt models combining different explanatory variables, training methods, ARIMA parameters, and window sizes, one may want to test statistically the significance of certain functional classes within the selection over certain periods. Here we provide two approaches: the simple Bayesian framework where we set the prior to be proportional to the inverse of the size of the hypothesis class and updates the likelihood by simple counting, and a frequentist framework whereby binomial distribution can be assumed and thus p-values can be produced.

\subsection{Method}
\subsubsection{A simple Bayesian hypothesis testing}
For a subset of functions $H^1 \subsetneq H^0 \subseteq H$, consider the following hypothesis:
\begin{align*}
H_0:& \text{Functions in the set $H^1$  have the same or lower chance of being selected than the ones in $H^0 \setminus H^1$.} \\
H_A:& \text{Functions in the set $H^1$  have a higher chance of being selected than the ones in $H^0 \setminus H^1$.}
\end{align*}

Note that here, "being selected" refers to that a particular function in the functional set $H^0$ or $H^1$ being selected by the adaptive learning model.

We consider a Bayesian framework to test the above hypothesis: write $\pi(H_0)$ and $\pi(H_A)$ as the prior for $H_0$ and $H_A$, respectively, and $\pi(y_t|H_0)$, $\pi(y_t|H_A)$ as the (conditional) likelihood for a testing observation value $y_t$ based on the hypothesis $H_0$ and $H_A$ respectively.

We define the following probabilities: \begin{align}
\pi(H_0)&:= \frac{ | H^1 |  }{| H^0 |} \\
\pi(y_t | H_0) &:=  \mathds{1} [ \text{Function in } H^0 \setminus H^1 \text{ is selected} ] = \mathds{1} [h_t^* \in H^0 \setminus H^1] \\
\pi(y_t | H_A) &:=  \mathds{1} [ \text{Function in } H^1 \text{ is selected} ] = \mathds{1} [ h_t^* \in H^1]
\end{align}
And as a result, given a period of data $\vy:=\{y_t\}_{t=T_0}^{T_1}$, the Bayes factor for $H_A$ is computed as:
\begin{subequations}
	\begin{align}
B_{A0}(\vy) &:= \frac{\pi(H_A | \vy)}{\pi(H_0 | \vy)} \\
&= \frac{\pi(H_A) \times \pi(\vy|H_A) }{\pi(H_0) \times \pi(\vy|H_0)} \\
&= \frac{| H^0 \setminus H^1 |}{ | H^1 |  } \times \frac{\sum_{t=T_0}^{T_1} \mathds{1} [ h_t^* \in H^1]}{\sum_{t=T_0}^{T_1}\mathds{1} [h_t^* \in H^0 \setminus H^1] } \label{20c}
\end{align}
\end{subequations}
We note from this setting, that if the true underlying process was $H_0$, the Bayes factor  tends to 1 or lower, and otherwise higher than 1. In case the fraction in \autoref{20c} contains 0 in the denominator, we assign the Bayes factor with infinity and assign a high value in the plot for illustration.

\subsubsection{A  frequentist hypothesis testing}

%We inherit the same hypothesis as stated in the previous subsection.
Another way to test the hypothesis, or a more common frequentist approach, can be done  by simply assuming, in case of $H_0$, a model in $H^1$ is chosen with probability $\frac{|H^1|}{|H^0|}$ as we assume no better performance under the null.

Suppose we have a period of test data $\vy:=\{y_t\}_{t=T_0}^{T_1}$, then write $n_1$ as the number of $t$ such that $h_t^*\in H^1$.\footnote{That is, $n_1=\sum_{t=T_0}^{T_1} \mathds{1} [ h_t^* \in H^1]$.} Under null hypothesis, we have $n_1\sim Bin(T_1-T_0+1 ,  \frac{ | H^1 |  }{| H^0 |} )$ as there are $T_1-T_0+1$ number of observations each with a likelihood of at most $\frac{ | H^1 |  }{| H^0 |}$ to be chosen.\footnote{More precisely, here we follow the traditional step to obtain a Binomial distribution from $T_1-T_0+1$ number of iid Bernoulli with probability being the set as the proportion.} We can therefore set the relevant critical value for the hypothesis testing, as well as the p-value. As usual, a close-to-zero (usually set as less than 0.05) p-value indicates a rejection of null in favour of the alternative $H_A$.

\subsection{Result}
We use the above approach to test seven hypotheses, each over a period of five trading days, within which there are 180 forecasting samples.\footnote{When required later, a higher-frequency can also be made, for which the results are more spiky due to the small sample.} We use three adaptive learning models to test simultaneously: we label $M_1$ as the one with $\lambda=1$ from model group 13, $M_2$ as the one with $\lambda=0.8$ from model group 13, and $M_3$ as the one with $\lambda=1$ from model group 14 type 1. All of them are with $(C_1,C_2)=(50\%,75\%)$. The choices of $M_1$ and $M_3$ are because of their good performance in MSE (see \autoref{tab6}) while $M_2$ is picked as a representative of low $\lambda$ --- which corresponds to a shorter-memory selection, aiding statistical conclusion here, though perform relatively badly in MSE ranking. Also, it is worth taking note that $M_3$ may be slightly questionable while performing the frequentist test, as the set-ups from the model group 14 could lead to highly-correlated model choices due to the term $D(h,h^*_{t-1})$, which contradicts with the underlying assumption of binomial distribution in the null hypothesis.

\begin{figure}[H]
	\centering
	\includegraphics[width=0.5\linewidth]{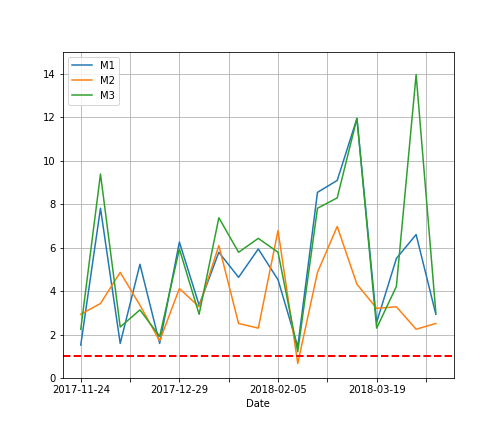}\hspace{-0.5cm}
	\includegraphics[width=0.5\linewidth]{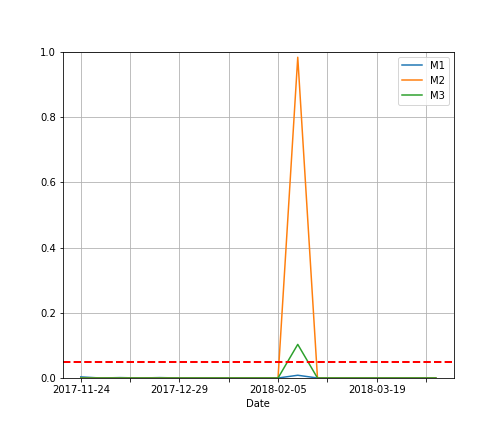}
	\caption{Bayesian testing result (left) and frequentist testing result (right)}
	\label{fig:fig1-2}
\end{figure}

The first hypothesis testing is to test the window size: whether it is 96 or not. Hence the functional set is $H^1=\{h(\iota, w, p,d,q)| w=96 \}$ and $H^0=H$. As shown in \autoref{fig:fig1-2}, we see this to be significant on many days for both tests --- a high Bayes factor can be observed in all models, for all periods but one. Likewise, a close to, if not 0, p-value can also be observed for a almost all periods. Such a frequent rejection of the null hypothesis means that the adaptive learning models still have a high reliance on the large-windowed models.

\begin{figure}[H]
	\centering
	\includegraphics[width=0.5\linewidth]{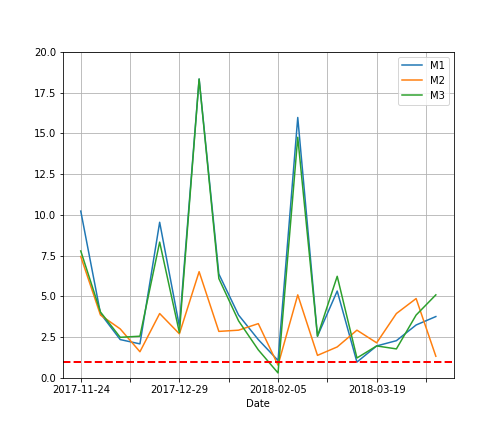}\hspace{-0.5cm}
\includegraphics[width=0.5\linewidth]{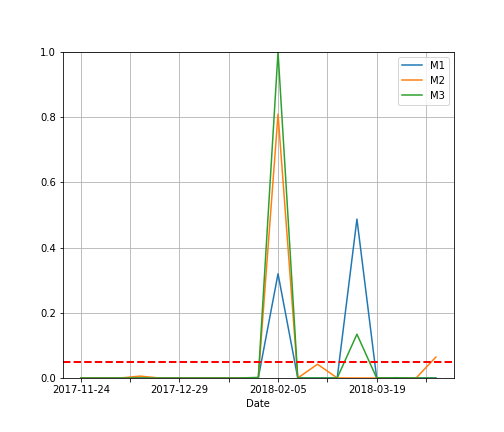}
\caption{Bayesian testing result (left) and frequentist testing result (right)}
	\label{fig:fig2-2}
\end{figure}

We now consider a hypothesis about whether the model group is zero. Hence the functional set is $H^1=\{h(\iota, w, p,d,q)| \iota=0 \}$ and $H^0=H$. We see from \autoref{fig:fig2-2} that the significance is high in most periods. Hence we could conclude these adaptive learning models to have a high occupation of model group zero, therefore the use of explanatory variable could be low, in many periods.

\begin{figure}[h]
	\centering
	\includegraphics[width=0.5\linewidth]{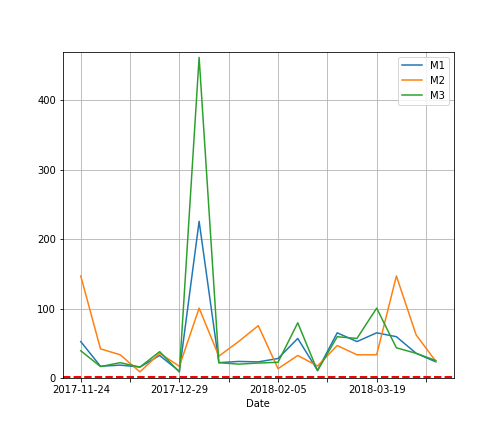}\hspace{-0.5cm}
	\includegraphics[width=0.5\linewidth]{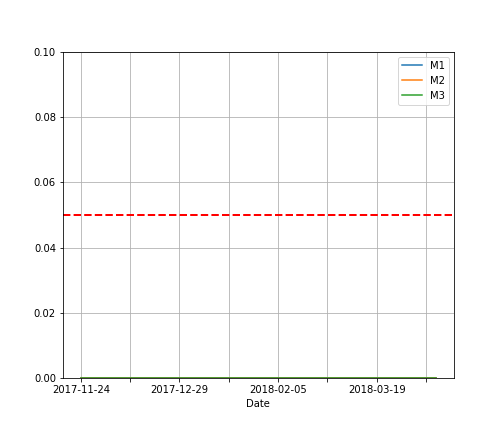}
	\caption{Bayesian testing result (left) and frequentist testing result (right)}
	\label{fig:fig3-2}
\end{figure}

We question about the training method --- whether the parameters are trained in univariate or vector models. Here we have $H^1=\{h(\iota, w, p,d,q)| \iota \leq 6 \}$ and $H^0=H$. From the previous examples, the univariate choice may be intuitively true, as multivariate models are rarely used, and from the testing results, such an intuition is confirmed. Indeed, as shown in \autoref{fig:fig3-2}, in both Bayesian and the frequentist, significance can be shown in all periods, and the magnitude of the Bayes factor is also gigantic.

\begin{figure}[h]
	\centering
	\includegraphics[width=0.5\linewidth]{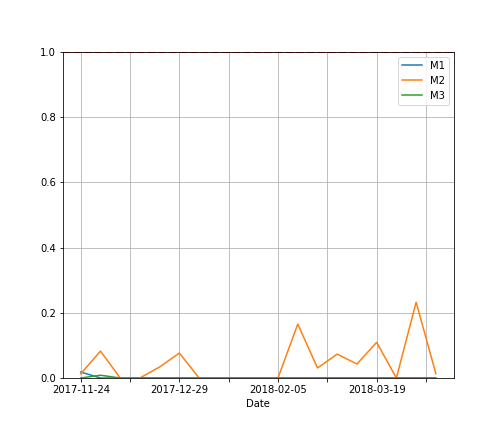}\hspace{-0.5cm}
	\includegraphics[width=0.5\linewidth]{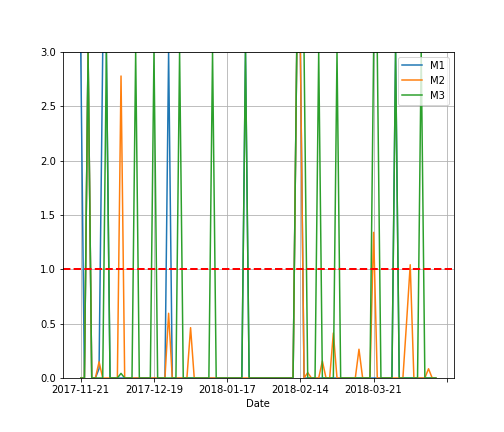}
	\caption{Bayesian testing result in a 5-day period (left) and in an 1-day period (right)\\ Note on the right panel: infinity is plotted as 3 here.}
	\label{fig:fig4-2}
\end{figure}

Now, we wonder if the significance of small-window model exists, compared to the already-tested significant large-window models (from \autoref{fig:fig1-2}). In this case, we have $H^1=\{h(\iota, w, p,d,q)| w=12 \}$ and $H^0=\{h(\iota, w, p,d,q)| w\in \{12,96\} \}$. The p-values are almost all 1, and the Bayes factors, as shown in the left panel of \autoref{fig:fig4-2}, is rather low, meaning the small-windows model are not picked significantly compared to the large one. We also attempt to "zoom-in" by using a 1-day period, for which the Bayes factors are plotted in the right panel. The p-values are mostly 1 or close-to 1, while the Bayes factors, as can be seen, are somewhat spiky --- this can be due to the way the Bayes factor is constructed, but it suggests some instability as the value is spiky, for most models. This means, despite the small-window models are non-significant at a larger scale, they may occasionally play a part, as suggested by the Bayes factors on a 1-day period.

\begin{figure}[H]
	\centering
	\includegraphics[width=0.5\linewidth]{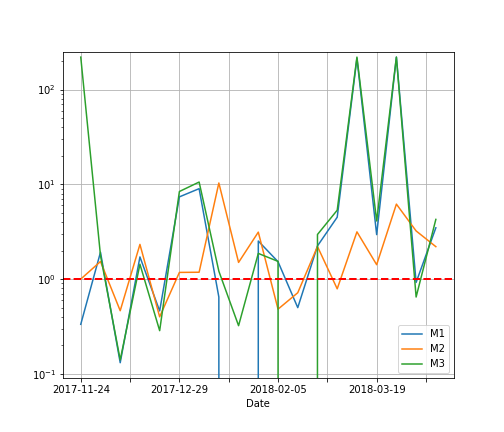}\hspace{-0.5cm}
	\includegraphics[width=0.5\linewidth]{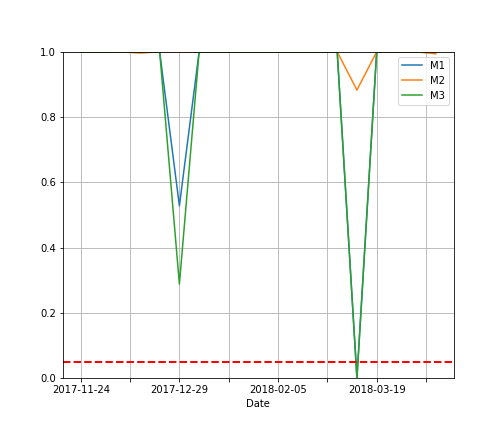}
	\caption{Bayesian testing result (left) and frequentist testing result (right)\\ Note on the left panel: infinity is plotted as 220 here, with a logarithm scale $(\log_{10})$ being used.}
	\label{fig:fig5-2}
\end{figure}

Likewise, in the followings we test the significance of certain choices of variables compared to a wider set, in particular, we first consider the sole choice of OFI mean compared against its combination with OIB mean or solely the OIB mean, which leads to $H^1=\{h(\iota, w, p,d,q)| \iota \in \{2,8\} \}$ and $H^0=\{h(\iota, w, p,d,q)| \iota \in \{1,2,3,7,8,9\} \}$. Test results, as shown in \autoref{fig:fig5-2}, suggest no dominance by the OFI mean for all but one period from a frequentist viewpoint, while the Bayes factor is occasionally large, meaning significance may exist for some periods. This is due to the fact that the  set by null hypothesis $(H^0)$ may not occupy a large amount of functions being chosen,\footnote{As shown in \autoref{fig:fig2-2}, the model without explanatory variable is of high significance.} contributing to a spiky and potentially large Bayes factor by occasion.

We now run the same test for OFI p-score compared against its combination with OFI p-score or solely the OIB p-score. Hence $H^1=\{h(\iota, w, p,d,q)| \iota \in \{5,11\} \}$ and $H^0=\{h(\iota, w, p,d,q)| \iota \in \{4,5,6,10,11,12\} \}$.  As shown in \autoref{fig:fig6-2} in the appendix, the significance is not huge although the Bayes factor may be occasionally high --- this, as explained in the previous part, can be purely due to the lack of samples occupied by $H^0$.

\begin{figure}[H]
	\centering
	\includegraphics[width=0.5\linewidth]{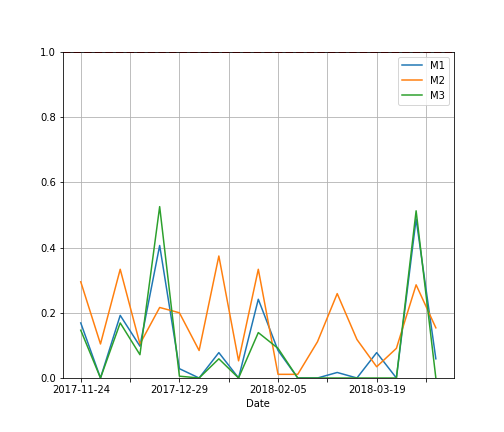}\hspace{-0.5cm}
	\includegraphics[width=0.5\linewidth]{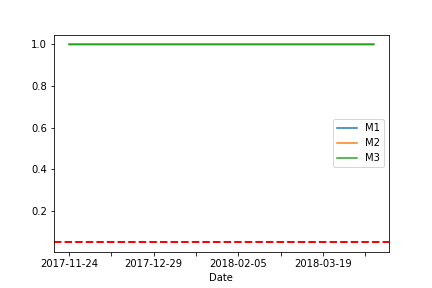}
		\includegraphics[width=0.5\linewidth]{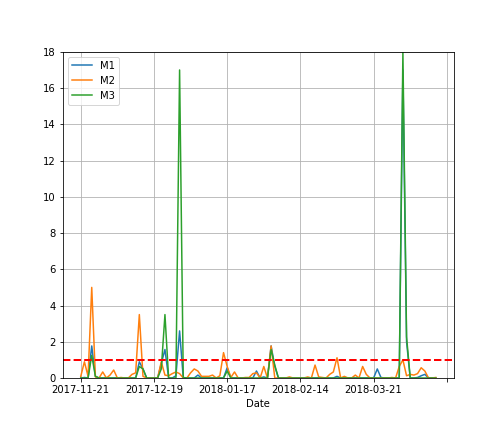}\hspace{-0.5cm}
	\includegraphics[width=0.5\linewidth]{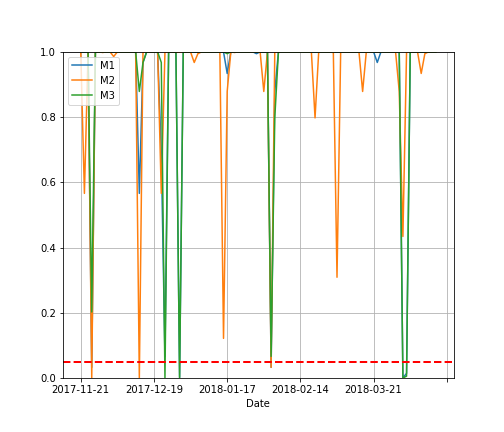}
	\caption{Bayesian testing result (left) and frequentist testing result (right) with 5-day period (up) and 1-day period (down)\\ Note on the lower left panel: infinity is plotted as 18 here.}
	\label{fig:fig7}
\end{figure}

Finally, we concern about the choice of the difference order, thus $H^1=\{h(\iota, w, p,d,q)| d=2 \}$ and $H^0=H$. As shown in \autoref{fig:fig7}, while testing this in a 5-day period, one may easily conclude the insignificance of $H^1$ due to its consistently low Bayes factor and likewise the constantly almost-unity in the p-value; once zoomed into the 1-day period, the result becomes spiky in both the Bayesian and the frequentist tests, implying that the significance of 2nd difference being chosen is occasionally high.

As a conclusion from these testing, we see a significant component of large-window sized models and models without explanatory variables being selected by the top learning models, while instability and spikiness occur, in every aspect of the model choices, especially as we zoom into a 1-day period. This is because occasionally, statistical significance can be found for 2nd difference, small-window models, as well as groups of some explanatory variables.

%\subsection{A study on chaotic markets and trading curbs}

\pagebreak

\section{Discussion, Extension, and Conclusion}\label{S5}
As a general discussion about the adaptive learning, we reflect first from the statistical intuition: what is $h_t$ and why do we care to learn $h_t$? The ever-changing structure, as being frequently studied, requires certain awareness by the model on the time-variability, not only on the parameters, but also the functional forms. Classical treatments on time series, e.g.
\textcites{H1}{H2} offer the chance of parameter variation, while modern deep learning models, e.g. \textcite{SirignanoCont} can help on learning the $h_t$.

However, some issue may occur in the learning on $h_t$ --- interpretability is one, and ultimately the design could be questionable. Here we design a model-adapting and time-adaptive learning regime, to offer higher interpretability and  allow testing to be undertaken. Statistical conclusions may also be drawn from the adaptive learning models, for example, the explanatory variables may not be of good use for many occasions, as concluded in the testing. 

The design, particularly \autoref{AL2} and \autoref{AL4}, allows $H$ to be potentially infinite. For instance, one could set $H$ to contain infinitely many $p$ but set the move to be at most 1 from each time, i.e. $D(h,h_{t-1}^*)$ takes infinity if $|p-p_{t-1}^*|>1$. This allows the model to be theoretically more variable and contributes to the ultimate learning on $h_t$. Another more computationally expensive, but also important extension is to dynamically learn the "hyper-parameter". Here we have a variety of parameters, $C_1(t),C_2(t)$ for example, being set with constant proportion to certain statistics from the past --- these constants are which the model could have learnt, though the search of which would take high computational power. An interesting extension would be to learn these parameters and discuss the improvement on the learning. 

We also note the financial applicability of such an adaptive learning model: extra parameters can also be introduced to engage with practical application to trading, e.g. the loss function in \autoref{AL2} could take a specification that relates to a rolling-averaged profit and loss, or a more realistic profit and loss with trading barriers, e.g. $\sum_{t=t_s+1}^{t_s+17} sign(\alpha_t)\mathds{1}[ \alpha_t \in R(\Phi_t,H)] \frac{P_{t+1}-P_t}{P_t}$ where $R(\Phi_t,H)$ is a time-varying region for the signal to be strong enough to trade, which can be one of the parameters being learnt. Such an extension may also engage with the contemporary econometric methods in conditional heteroskedasticity \parencite{ACH2013}.

An additional direction is to engage with the study on penalisation, in particular the penalisation on estimation. In time series, due to the moving average terms, MLE is mostly inevitable, thus the penalisation must be done in the fashion of penalised MLE (pMLE). Here we focus on the penalisation on functional forms post-estimation while using the MLE at the first stage. An extension would be to engage with the theory of pMLE, e.g. from \textcites{CCG}{Spokoiny2018}, and use the adaptive learning to appreciate the value of penalisation in the context of forecasting. 
A particular modelling issue, as has been shown here, is the distaste towards multivariate models --- this can be due to the failure of capturing the underlying time-varying parameters together with low degrees of freedom. This has also been studied by  \textcite{WBBM} and more thoughts on penalisation and potentially "smart identification" using the past information could be worked on.

In conclusion, we propose a forecast-centric learning model that aims to adapt to the past information in a time series context. Such a model requires inputs of different functional forms, as well as training methods to produce parametric estimation and forecasting --- these are handled by the traditional ARIMA models and explanatory variables which are generated from the order book. The result of the learning model is comparable to the top models if the models were to be fixed, and can outperform the fixed models in relatively volatile and non-stationary market conditions. Additionally, stability is more ensured and the error-handling, functional penalisation, and potentially other criteria can be encoded as part of the model selection process. Since the process is intuitive and interpretable, many extensions and applications can be made, for which we have shown an application to statistical testing, in both a Bayesian and a frequentist context.

%This could also be explained by the fact that the order book data may are not as desirable as those ones from the American exchanges --- this can be apparent from the exploratory data analysis (\autoref{fig:2-2}) where the pure features could perform badly and correlate to the generic buy-and-hold strategy in terms of SR statistic.

%"hyper-parameter" learning

%Financial application --- not that miraculous results purely from order book

%Penalisation engagement

%\pagebreak

\appendix
%\section{Mathematics, Statistics, and Algorithms}
%\subsection{Data Cleaning: method for the VWM computation}
%\subsection{Details for implementing an ADF test}

%p=2 constant but no trend
%\subsection{Clarification on the set of functions}

\pagebreak
%\begin{comment}

\section{Additional statistical notes}
\subsection{On the ADF test}\label{ADF}
Here we adapt the following procedure for ADF Test with a constant. Let $\{y_t\}_{t=T_0}^{T_1}$ be the dataset we would like to run the test on.  Regress \begin{equation}\label{ADFeq}
y_t = \mu + \sum_{j=1}^{p}\phi_j y_{t-j} + \varepsilon_t, \hspace{0.5cm} \varepsilon_t \sim iid N (0,\sigma^2), \hspace{0.5cm} t\in \{T_0+p,...,T_1\}
\end{equation}
The specific test is
\begin{align*}
H_0: & \ \phi_1=1 \\
H_1: & \ \phi_1 <1
\end{align*}
Then construct test statistics $t:=\frac{\hat{\phi_1}-1}{sd(\hat{\phi_1})}$ where $\hat{\phi_1}$ and $sd(\hat{\phi_1})$ are the MLE estimates and standard deviation estimates for  $\phi_1$ respectively from estimating \autoref{ADFeq}. The asymptotic result suggests\footnote{See \textcite{TS3}.}
$t \distas{H_0} DF_{T_1-T_0-p}$. Thus, at a picked significance level,  the critical value $t^c$ is picked from $DF_{T_1-T_0-p}$ distribution and conclude:\footnote{Note: $t^c$ is negative here.} \begin{itemize}
	\item Reject $H_0$ and conclude stationarity if $t<t^c$;
	\item Do not reject $H_0$ and conclude non-stationarity if $t\geq t^c$.
\end{itemize} 

There are a variety of choices of $p$ one could take, usually $p \leq (T_1-T_0+1)^\frac{1}{3}$ \parencite{DF1}. Here, to ensure consistency across different windows, we decide to put $p=2$, as otherwise the small-window tests and later models get over-fitted.

\subsection{On MAE and MSE}\label{MSE}
Let $\{y_{t+1|t}\}_{t \in T }$ be the forecasts and let $\{y_{t+1}\}_{t\in T}$ be the validation data. Write $card(T)$ as the cardinality of the index set $T$, then the MAE and MSE are:
 \begin{align*}
MAE=& (card(T))^{-1} \sum_{t \in T} |y_{t+1|t}-y_{t+1}| \\
MSE=& (card(T))^{-1} \sum_{t \in T} |y_{t+1|t}-y_{t+1}|^2
\end{align*}

\pagebreak
%\end{comment}

\section{Additional figures and tables}
\begin{table}[H]
	\centering
	\begin{tabular}{lcccccc}
	\toprule
	{} &  OFI mean &  OFI p-score &  OIBmean &  OIB p-score &     VWM  &  Number per bracket \\
	\midrule
	count &  5040 &     5040 &  5040 &     5040 & 5040 & 5040 \\
	mean  &     0.17 &        0.54 &     0.00 &        0.51 & 4068.21 & 424.79\\
	std   &     0.10 &        0.02 &     0.06 &        0.06 &  141.09 & 84.88\\
	min   &    -0.25 &        0.45 &    -0.39 &        0.20 & 3734.99 & 81\\
	25\% quantile  &     0.10 &        0.52 &    -0.03 &        0.47 & 3996.03 &377 \\
	50\% quantile  &     0.16 &        0.54 &     0.00 &        0.51 & 4052.03 &431\\
	75\%  quantile &     0.23 &        0.55 &     0.04 &        0.55 & 4137.43 &483\\
	max   &     0.95 &        0.62 &     0.33 &        0.77 & 4421.84 &601\\
	\bottomrule
\end{tabular}
\caption{Summary of statistics for features (column 1 to 4), the VWM (column 5), and the number of observations per 5-minute bracket (column 6).}
\label{Tab2.1}
\end{table}

\begin{table}[H]
	\centering

\begin{tabular}{|c|c|c|c|c|c|}
	\hline 
	Model Group & 0 & 1,3,7,9 & 2,3,8,9 & 4,6,10,12 & 5,6,11,12 \\ 
	\hline 
	Explanatory Variables & None & OIBmean & OFI mean & OIBp-score & OFI p-score \\ 
	\hline 
\end{tabular} 

\caption{Explanatory variables used in each of the fixed model groups}
\label{TabB.1}
\end{table}

\begin{table}[H]
	\centering
	\begin{tabular}{|c|c|c|c|c|c|c|}
		\hline 
		$\lambda$ & 1 & 0.99 & 0.95 & 0.9 & 0.85 & 0.8 \\ 
		\hline 
		$\log_\lambda(0.5)$ & $\infty$ & 68.97 & 13.51 & 6.58 & 4.27 & 3.11 \\ 
		\hline 
		$\lambda^{47}$ & 1 & 0.62 & 0.09 & 0.01 & 0 &0  \\ 
		\hline 
	\end{tabular} 
\caption{Logarithmic and exponential properties amongst the choices of $\lambda$ (Up to two decimal places)}
\label{TabB.2}
\end{table}
\pagebreak
\begin{figure}[H]
	\centering
	\includegraphics[width=\linewidth]{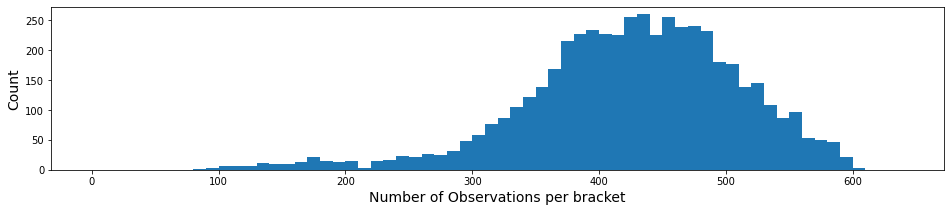}
	\includegraphics[width=\linewidth]{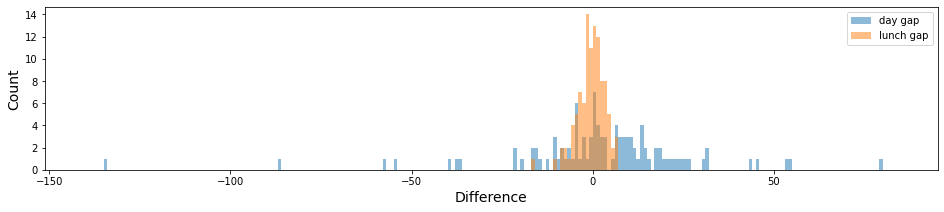}
	\includegraphics[width=\linewidth]{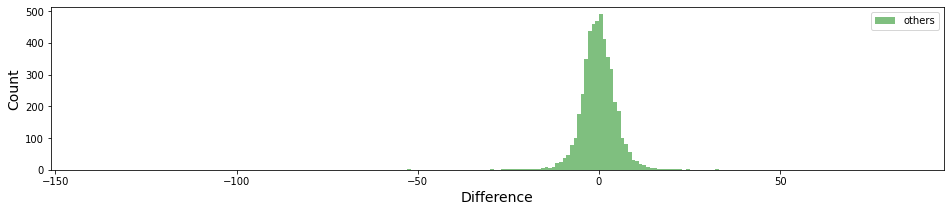}
	\caption{Upper: histogram of the number of observations per bracket. \\ Centre and Lower: histogram of the day gap, lunch gap, and the others, laid down on the same scale.}
	\label{fig:B1}
\end{figure}

\begin{figure}[H]
	\centering
	\includegraphics[width=0.5\linewidth]{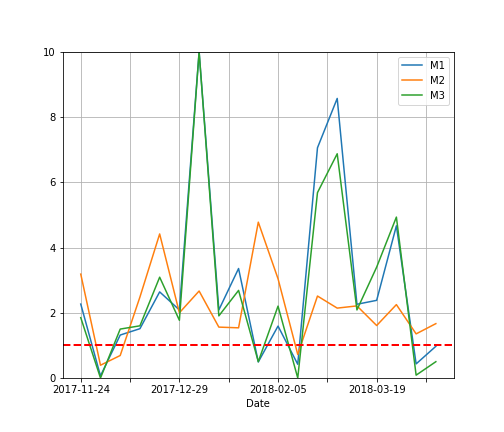}\hspace{-0.5cm}
	\includegraphics[width=0.5\linewidth]{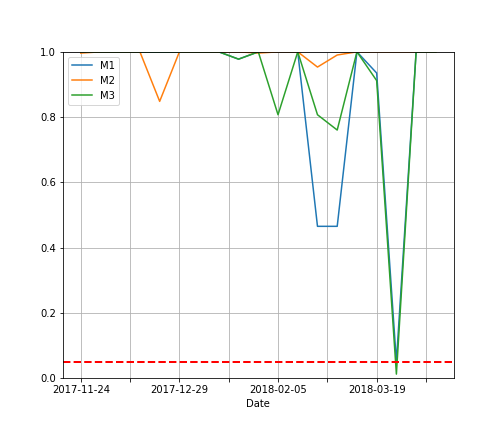}
	\caption{Bayesian testing result (left) and frequentist testing result (right)}
	\label{fig:fig6-2}
\end{figure}

\begin{comment}
\pagebreak
\section{Codes}
This section of the appendix, as requested by the dissertation project section (subsection 3.5) of the MSc Handbook Version 1.3, contains the essential codes for implementing the data cleaning, modellings, and most of the visualisations and table productions.

In what follows, we attach, notebooks by notebooks, sections of codes that relates to each of the titled purposes of the notebooks. Some \texttt{.py} scripts are produced as pure functions or library collections, which are implemented by executing them in the notebooks.

\pagebreak
\includepdf[pages=-,width=\textwidth]{../Codebook/Data/Tools.pdf}
\includepdf[pages=-,width=\textwidth]{../Codebook/Data/Data.pdf}
\includepdf[pages=-,width=\textwidth]{../Codebook/FM/UM.pdf}
\includepdf[pages=-,width=\textwidth]{../Codebook/FM/FM.pdf}
\includepdf[pages=-,width=\textwidth]{../Codebook/LM/LM.pdf}
\includepdf[pages=-,width=\textwidth]{../Codebook/MS/MS.pdf}
\includepdf[pages=-,width=\textwidth]{../Codebook/tests/tests.pdf}
\pagebreak
\end{comment}
%\section{References}
\printbibliography
%\section{Codes (for dissertation ONLY)}
\end{document}